\documentclass[11pt]{article}

\usepackage{graphicx}
\usepackage{amsmath}
\usepackage{amsfonts}
\usepackage{amssymb}
\usepackage{verbatim}

\renewcommand{\arraystretch}{1.3}
\catcode`\@=11
\def\marginnote#1{}

\newcount\hour
\newcount\minute
\newtoks\amorpm
\hour=\time\divide\hour by60
\minute=\time{\multiply\hour by60 \global\advance\minute by-\hour}
\edef\standardtime{{\ifnum\hour<12 \global\amorpm={am}%
        \else\global\amorpm={pm}\advance\hour by-12 \fi
        \ifnum\hour=0 \hour=12 \fi
        \number\hour:\ifnum\minute<10 0\fi\number\minute\the\amorpm}}
\edef\militarytime{\number\hour:\ifnum\minute<10 0\fi\number\minute}

\def\draftlabel#1{{\@bsphack\if@filesw {\let\thepage\relax
      \xdef\@gtempa{\write\@auxout{\string
          \newlabel{#1}{{\@currentlabel}{\thepage}}}}}\@gtempa \if@nobreak
    \ifvmode\nobreak\fi\fi\fi\@esphack} \gdef\@eqnlabel{#1}}
    \def\@eqnlabel{}
\def\@vacuum{}
\def\draftmarginnote#1{\marginpar{\raggedright\scriptsize\tt#1}}

\def\draft{
  \oddsidemargin -.5truein
  \def\@oddfoot{\footnotesize \sl preliminary draft \hfil
    \rm\thepage\hfil\sl\today\quad\militarytime}
  \let\@evenfoot\@oddfoot \overfullrule 3pt
    \let\label=\draftlabel
    \let\marginnote=\draftmarginnote
  \def\@eqnnum{(\theequation)\rlap{\kern\marginparsep\tt\@eqnlabel}%
    \global\let\@eqnlabel\@vacuum}

  }
\makeatletter
\newdimen\normalarrayskip              % skip between lines
\newdimen\minarrayskip                 % minimal skip between lines
\normalarrayskip\baselineskip
\minarrayskip\jot
\newif\ifold             \oldtrue            \def\new{\oldfalse}
\def\arraymode{\ifold\relax\else\displaystyle\fi} % mode of array entries
\def\eqnumphantom{\phantom{(\theequation)}}     % right phantom in eqnarray
\def\@arrayskip{\ifold\baselineskip\z@\lineskip\z@
     \else
     \baselineskip\minarrayskip\lineskip2\minarrayskip\fi}
\def\@arrayclassz{\ifcase \@lastchclass \@acolampacol \or
\@ampacol \or \or \or \@addamp \or
   \@acolampacol \or \@firstampfalse \@acol \fi
\edef\@preamble{\@preamble
  \ifcase \@chnum
     \hfil$\relax\arraymode\@sharp$\hfil
     \or $\relax\arraymode\@sharp$\hfil
     \or \hfil$\relax\arraymode\@sharp$\fi}}
\def\@array[#1]#2{\setbox\@arstrutbox=\hbox{\vrule
     height\arraystretch \ht\strutbox
     depth\arraystretch \dp\strutbox
     width\z@}\@mkpream{#2}\edef\@preamble{\halign
\noexpand\@halignto
\bgroup \tabskip\z@ \@arstrut \@preamble \tabskip\z@ \cr}%
\let\@startpbox\@@startpbox \let\@endpbox\@@endpbox
  \if #1t\vtop \else \if#1b\vbox \else \vcenter \fi\fi
  \bgroup \let\par\relax
  \let\@sharp##\let\protect\relax
  \@arrayskip\@preamble}
\def\eqnarray{\stepcounter{equation}%
              \let\@currentlabel=\theequation
              \global\@eqnswtrue
              \global\@eqcnt\z@
              \tabskip\@centering
              \let\\=\@eqncr

 \halign to \displaywidth\bgroup
    \eqnumphantom\@eqnsel\hskip\@centering
    $\displaystyle \tabskip\z@ {##}$%
    \global\@eqcnt\@ne \hskip 2\arraycolsep
         $\displaystyle\arraymode{##}$\hfil
    \global\@eqcnt\tw@ \hskip 2\arraycolsep
         $\displaystyle\tabskip\z@{##}$\hfil
         \tabskip\@centering
    &{##}\tabskip\z@\cr}
\newfont{\hr}{msbm10}
\newfont{\ams}{msam10}
\hoffset= - 0.9in         % switch off for draft style
\def\beq{\begin{equation}}
\def\eeq{\end{equation}}
\def\ba{\beq\new\begin{array}{c}}
\def\ea{\end{array}\eeq}
\def\be{\ba}
\def\ee{\ea}
\def\stackreb#1#2{\mathrel{\mathop{#2}\limits_{#1}}}

\def\F{{\cal F}}

\def\d{\partial}
\def\N2{${\cal N}=2$}

\def\1N{${\cal N}=1$}
\def\4N{${\cal N}=4$}
\def\nn{\nonumber}
\def\half{{\textstyle{1\over2}}}

\def\p{\partial}

\def\la{\left\left<}
\def\ra{\right\right>}
\newcommand{\rf}[1]{(\ref{#1})}

\newdimen\linethick  \linethick=0.4pt
\newdimen\hboxitspace    \hboxitspace=5pt
\newdimen\vboxitspace    \vboxitspace=5pt

\def\fr#1{%
\beq\new
\vcenter{
\hrule height\linethick
          \hbox{\vrule width\linethick
                \kern\hboxitspace
                \vbox{\kern\vboxitspace
                      \hbox{$\begin{array}{c}\displaystyle#1
         \end{array}$}%
                      \kern\vboxitspace}%
                \kern\hboxitspace
                \vrule width\linethick}%
          \hrule height\linethick}%
\eeq}
\renewcommand{\d}{\partial}

\def\F{\mathcal{F}}

\renewcommand{\tt}[1][mer]{\hbox{\tiny{#1}}}

\newcommand{\Tr}{\mathop{\rm Tr}\nolimits}

\def\p{\partial}
\def\tr{{\rm tr}\,}
\def\Tr{{\rm Tr}\,}

\def\p{\partial}
\def\tr{{\rm tr}\,}
\def\Tr{{\rm Tr}\,}

\def\la{\left<}
\def\ra{\right>}
\def\lala{\la\!\!\!\!\la}
\def\rara{\ra\!\!\!\!\ra}
\def\bla{\Big<}
\def\bra{\Big>}
\def\blabla{\bla\!\!\bla}
\def\brabra{\bra\!\!\bra}

\def\BB{{1/2b}}

\hoffset= - 1.0in         % switch off for draft style
\title{{\bf
On AGT Relations with Surface Operator Insertion and Stationary
Limit of Beta-Ensembles} %\vspace{.1cm}
}
\author{{\bf A. Marshakov}\thanks{E-mail: \ mars@itep.ru; mars@lpi.ru}\ ,\ \
\date{ } %\\ {\small
{\bf A. Mironov}\thanks{E-mail:
\ mironov@itep.ru; mironov@lpi.ru}
\date{ } \\ \bigskip
{\small {\it Theory Department, Lebedev Physics Institute}
and {\it ITEP, Moscow, Russia}}\\
{\bf A.Morozov}\thanks{E-mail: \ morozov@itep.ru}
\date{ } \\ {\small
{\it ITEP, Moscow, Russia}}}

\begin{document}

\setcounter{footnote}{3}

\setcounter{tocdepth}{3}

\maketitle

\vspace{-8.cm}

\begin{center}
\hfill FIAN/TD-07/10\\
\hfill ITEP/TH-38/10 %\\
%\hfill MPIM-???
\end{center}

\vspace{5.cm}

\begin{abstract}
\noindent
We present a summary of current knowledge about the AGT relations for
conformal blocks with additional insertion of
the simplest degenerate operator, %$V_{1/2b}(x)$
and a special choice of the corresponding intermediate dimension,
when the conformal
blocks satisfy hypergeometric-type differential equations in position %$x$
of the degenerate operator. A
special attention is devoted to representation of conformal block
through the beta-ensemble resolvents and to its asymptotics in the
limit of large dimensions (both external and intermediate) taken asymmetrically
in terms of the deformation epsilon-parameters.
The next-to-leading term in the
asymptotics defines the generating differential %$dS(x;\epsilon_1)$
in the Bohr-Sommerfeld representation of the one-parameter deformed Seiberg-Witten
prepotentials (whose full two-parameter deformation leads to Nekrasov functions).
This generating differential is also shown to be the one-parameter version
of the single-point resolvent for the corresponding
beta-ensemble, and its periods in the perturbative limit of the
gauge theory are expressed through the ratios of the Harish-Chandra function.
The Shr\"odinger/Baxter equations, considered earlier in this context,
directly follow from the differential
equations for the degenerate conformal block. This provides a powerful method
for evaluation of the
single-deformed prepotentials, and even for the Seiberg-Witten
prepotentials themselves. We mostly concentrate on the
representative case of the insertion into the four-point block on sphere
and one-point block on torus.
\end{abstract}

\section{Introduction}

The AGT conjecture \cite{AGT}
establishes explicit relations between
the basic formulas in several principal branches
of modern theory and naturally attracts an increasing
attention \cite{AGT}-\cite{Yam}.
The main objects of investigation are various conformal
blocks \cite{CFT}, and the statement is that they can
be also represented

\noindent
(1) as matrix model \cite{confmamo} and/or beta-ensemble \cite{beta}
partition functions in the Dijkgraaf-Vafa (DV) phase \cite{DVmamo,DVph,DVnext},

\noindent
(2) as LMNS integrals \cite{LMNS},

\noindent
(3) as combinations of the Nekrasov functions \cite{NF}
(i.e. as a generalization of hypergeometric series expansions,
\cite{MMnf}),

\noindent
(4) as exponentials of the deformed or ``quantized''  \cite{MMbz}
Seiberg-Witten (SW) prepotentials \cite{SW},
described in terms of integrable systems \cite{SW1,intSWd,intSW},
and so on.

\noindent
The AGT relations reflect a duality pattern \cite{dua},
associated with the twisted compactification of
the non-Lagrangian superconformal $6d$ theory \cite{6dth}
for a $M5$ brane
on a two-dimensional Riemann surface with boundaries,
giving rise to a four-dimensional ${\cal N}=2$ supersymmetric Yang-Mills theory,
which can be further compactified down to 3,2,1,0 dimensions.

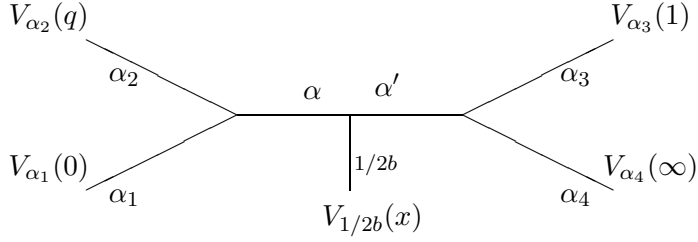
\begin{figure}
\unitlength 1mm % = 2.845pt
\linethickness{0.4pt}
\ifx\plotpoint\undefined\newsavebox{\plotpoint}\fi
\begin{picture}(100,40)(-60,-17)
\put(0,0){\line(1,0){30}}
\put(0,0){\line(-2,1){20}}
\put(0,0){\line(-2,-1){20}}
\put(30,0){\line(2,1){20}}
\put(30,0){\line(2,-1){20}}
\put(15,0){\line(0,-1){10}}
\put(-25,-7){\makebox(0,0)[cc]{$V_{\alpha_1}(0)$}}
\put(-25,13){\makebox(0,0)[cc]{$V_{\alpha_2}(q)$}}
\put(55,13){\makebox(0,0)[cc]{$V_{\alpha_3}(1)$}}
\put(55,-7){\makebox(0,0)[cc]{$V_{\alpha_4}(\infty)$}}
\put(18,-14){\makebox(0,0)[cc]{$V_{\BB}(x)$}}
%\put(-3,-18){\makebox(0,0)[cc]{$V_{\BB}(x)$}}
\put(-15,-11){\makebox(0,0)[cc]{$\alpha_1$}}
\put(-15,5){\makebox(0,0)[cc]{$\alpha_2$}}
%\put(-4,-5){\makebox(0,0)[cc]{$\alpha_{11}$}}
\put(18,-7){\makebox(0,0)[cc]{$\phantom.^{\BB}$}}
\put(45,5){\makebox(0,0)[cc]{$\alpha_3$}}
\put(45,-11){\makebox(0,0)[cc]{$\alpha_4$}}
%\put(15,3){\makebox(0,0)[cc]{$\alpha=a+\epsilon/2???$}}
\put(10,3){\makebox(0,0)[cc]{$\alpha$}}
\put(20,4){\makebox(0,0)[cc]{$\alpha'$}}
\end{picture}
\caption{{\footnotesize
Here $z_{1,2,3,4} = (0,q,1,\infty)$ and $q\ll x \ll 1$.
In conformal theory, the structure constant for degenerate
primary vanishes unless $\alpha' =\alpha \pm \BB$ \cite{CFT}.
In free field representation of \cite{MMS1,MMS}
for $\alpha' \neq \alpha \pm \BB$, there are additional screening
insertions in the matrix model ($\beta$-ensemble) representation,
with {\it open} integration contours stretching from $0$ to $x$.
As explained in s.\ref{3.2.2} such insertions violate differential
equations naively following from the equation (\ref{degdifeq})
for the degenerate field, (\ref{degeq}).
Therefore, in this paper we consider only the case
of $\alpha'=\alpha \pm \BB$.
The relation to the $a$-parameter in Yang-Mills theory is
$\alpha = a+\epsilon/2$.
In the limit of $\epsilon_2\rightarrow 0$
the difference $1/2b = \frac{1}{2}\sqrt{-\frac{\epsilon_2}{\epsilon_1}}$
between $\alpha$ and $\alpha'$ gets negligible and $a$
in the corresponding Nekrasov function $F(\epsilon_1)$ at this limit can be
considered as related to either $\alpha$ or $\alpha'$,
thus restoring the symmetry of the diagram in application
to the AGT relation.
}}
\label{conf41}
\end{figure}

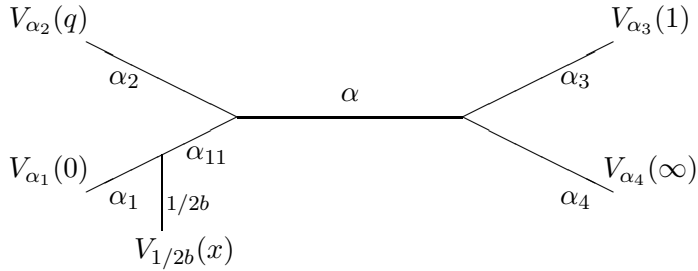
\begin{figure}
\unitlength 1mm % = 2.845pt
\linethickness{0.4pt}
\ifx\plotpoint\undefined\newsavebox{\plotpoint}\fi
\begin{picture}(100,40)(-60,-17)
\put(0,0){\line(1,0){30}}
\put(0,0){\line(-2,1){20}}
\put(0,0){\line(-2,-1){20}}
\put(30,0){\line(2,1){20}}
\put(30,0){\line(2,-1){20}}
\put(-10,-5){\line(0,-1){10}}
\put(-25,-7){\makebox(0,0)[cc]{$V_{\alpha_1}(0)$}}
\put(-25,13){\makebox(0,0)[cc]{$V_{\alpha_2}(q)$}}
\put(55,13){\makebox(0,0)[cc]{$V_{\alpha_3}(1)$}}
\put(55,-7){\makebox(0,0)[cc]{$V_{\alpha_4}(\infty)$}}
\put(-7,-18){\makebox(0,0)[cc]{$V_{\BB}(x)$}}
%\put(-3,-18){\makebox(0,0)[cc]{$V_{\BB}(x)$}}
\put(-15,-11){\makebox(0,0)[cc]{$\alpha_1$}}
\put(-15,5){\makebox(0,0)[cc]{$\alpha_2$}}
\put(-4,-5){\makebox(0,0)[cc]{$\alpha_{11}$}}
\put(-7,-12){\makebox(0,0)[cc]{$\phantom.^{\BB}$}}
\put(45,5){\makebox(0,0)[cc]{$\alpha_3$}}
\put(45,-11){\makebox(0,0)[cc]{$\alpha_4$}}
\put(15,3){\makebox(0,0)[cc]{$\alpha$}}
\end{picture}
\caption{{\footnotesize
Topology of the tree diagram implies certain ordering
of pairings in the definition of the conformal block.
From each OPE only the contribution of one particular Verma
module is picked up, thus, the associativity of OPE is restored
only after sums are taken over the intermediate dimensions. This diagram
corresponds to the ordering different from Fig.\ref{conf41}: $x\gg q\gg 1$.
Here the intermediate
dimension $\alpha_{11}=\alpha_1\pm1/2b$. The two diagrams are connected by a duality
transformation.
}}
\label{conf410}
\end{figure}

In this paper we review
the existing knowledge about these relations
in the particular case of the 4-point spherical conformal block with
the additional insertion of the simplest degenerate primary field
shown in Fig.\ref{conf41},
\be
B_5(x|z_i) = \ \left< V_{\BB}(x) \prod_{i=1}^4 V_{\alpha_i}(z_i)\right>
\label{Bdef}
\ee
as well as for the 1-point conformal block on a torus, and their degenerate limits.
For the spherical case we consider only this type of diagram, all other (e.g. that on Fig.2)
can be in principle obtained from that on Fig.1
by duality transformations (though we shall not consider this issue in the paper).

In the language of $4d$ SYM theory such insertion describes
a "surface operator", produced by $M2$-brane, which
lies entirely in the four-dimensional space-time
and is located at a point $x$ on the Riemann surface.
Within the CFT framework, this conformal block and the associated $5$-point
correlation function are the standards objects of interest
\cite{CFT,ZFLON}, since for a special choice of intermediate
dimension they satisfy the hypergeometric-type differential equations in $x$-variable,
which do {\em not} hold for generic conformal blocks.
This story has been already addressed in relation with the AGT conjecture in
\cite{Brav,AGTguk,FLit1,Wylsurf,MT}.

In this paper we describe the arrows in the following diagram:

$$
\begin{array}{ccccc}
&&\!\!\!\!\!\! \!\!\!\!\!\!B(x|z)&& \\
&&&&\\
&& \!\!\!\!\!\!\!\!\!\!\!\!\!\!\!\!\!
%\phantom.^{(4)}
\phantom.^{(\ref{degdifeq})}\swarrow\ \ \ \ \ \ \ \ \ \ \ \ \ \
\ \ \ \ \ \ \ \ \ \ \ \ \searrow^{(1)}&&\\
&&&&\\
\left({b^2}\p^2_x
- \sum_i \frac{\p_i}{x-z_i} -
\sum_i \frac{\Delta_i}{(x-z_i)^2}
\right)B(x|z) = 0
&&&& \!\!\!\!\!\!\!\!\!\!\!\!\!\!\!\!\!\!\!\!
B(x|z) = \ \blabla \det(x-M) \brabra \\
&&&&\downarrow \hbox{\scriptsize{(\ref{WyK5})}}\\
\ \ \ \ \ \ \downarrow \ \phantom.^{(4)}&&&&
\!\!\!\!\!\!\!\!\!\!\!\!\!\!\!\!\!\!\!\!\!\!\!\!\!\!\!\!\!\!\!\!\!\!\!\!\!\!\!\!
\log B(x|z) = \sum_{k}\frac{1}{k!}
\blabla \Big(\Tr\log(x-M)\Big)^k\brabra_{conn} \\
&&&&\\
&&&&\hspace{-6cm}\phantom.^{(\ref{76})}\swarrow\\
\log B(x|z) = \frac{F(\epsilon_1)}{\epsilon_1\epsilon_2} +
\frac{S(x;\epsilon_1)}{\epsilon_1} + O(\epsilon_2)
&&
\left\{ \begin{array}{c} a = \oint_A dS(x;\epsilon_1)
\\
\frac{\p F(\epsilon_1)}{\p a} = \oint_B dS(x;\epsilon_1)
\end{array} \right.
&&%\!\!\!\!\!\!\!\!\!\!\!\!\!\!\!\!\!\!\!\!\! \phantom.^{(\ref{76})}\swarrow
\\
\downarrow &&&& \\
\log B(x|z) = \frac{F_{SW}}{\hbar^2} +
\frac{S_{SW}(x)}{\hbar} + O(\hbar^0)
&&
\left\{ \begin{array}{c} a = \oint_A dS_{SW}(x)\\
\frac{\p F_{SW}}{\p a} = \oint_B dS_{SW}(x)
\end{array}\right.
&& \\
&&&&\\
\end{array}
$$

Hereafter $\bla\ldots\bra$ denote the CFT correlators, with
$\bla\ldots\bra_{free}$ stressing that this is the conformal theory of free
massless fields, while $\blabla\ldots\brabra$
denotes the $\beta$-ensemble averages, with the subscript $conn$ referring to the connected
correlators.

\bigskip

The right column deals with the matrix model (beta-ensemble) representation,
where the parameters of the conformal block define the shape of
the potential, the number of integrations (DV phase)
and the spectral complex curve.
In this approach, $B_5(x|z)$ can be  expressed
\cite{AGTguk,Wylsurf,Kanno} in terms of the exact resolvents of \cite{AMMEO},
which can be recursively constructed for any given spectral surface.

The left column makes use of the CFT equation for the null-vector
\be
\Big(b^2 L_{-1}^2- L_{-2}\Big)V_{\BB} = 0
\label{degeq}
\ee
or for the degenerate primary field $V_{\BB}(x)$.
This equation induces an equation for the conformal block $B_5(x|z)$
{\it only} provided the new intermediate dimension takes a
{\it special} value: such that the $\alpha$-parameters
of the two lines, attached to $V_{\BB}(x)$ (see Fig. 1) satisfy
\be
\alpha - \alpha' = \pm\BB,
\ee
see sect.~\ref{3.2.2} for details.
With this selection rule,
$B_5(x|z)$ satisfies the second-order differential equation, which
actually has a typical shape of a non-stationary Shr\"odinger equation
(cf. with \cite{Brav}; in fact, it takes literally the form of the
non-stationary Shr\"odinger equation
only in the specific limit of large dimensions which corresponds to the
pure gauge theory),
while the 4-point conformal block with the degenerate field satisfies
a stationary Schr\"odinger equation.

An important application of the $V_{1/2b}$ insertions into conformal blocks
is that they describe the $\epsilon_2\rightarrow 0$ limit
of the Nekrasov functions, we address to as {\it stationary} for the reasons
to be discussed below.
This limit is technically non-trivial and
very interesting, since it corresponds to a
quantization \cite{NS,MMbz} of the classical integrable
systems, associated to the supersymmetric gauge theories through
the standard dictionary of \cite{intSWd}\footnote{Under
this quantization, e.g. the spectral curve converts
into a Baxter equation.
For the further generalization of integrability,
provided by the double
deformation with both $\epsilon_1,\epsilon_2\neq 0$, see Conclusion.
}.
In general the SW representation of the conformal block,
\be
\frac{\p \log B_4(z)}{\p a_I} = b^2\oint_{B_I} \rho_1(x), \\
a_I = \oint_{A_I} \rho_1(x)
\ee
involves the exact one-point resolvent $\rho_1$ of the corresponding beta-ensemble
(Dotsenko-Fateev matrix model \cite{MMS1}),
which is a rather complicated quantity.
However, in the $\epsilon_2\rightarrow 0$ limit things get simplified,
in this limit the multi-trace correlators in the beta-ensemble
are factorized and the resolvent acquires a new representation
\be
\rho_1(x) = \blabla\tr \frac{1}{x-M}\brabra_{conn}
\ \stackrel{\epsilon_2=0}{=}\
\frac{\blabla\left(\tr \frac{1}{x-M}\right) \det (x-M)\brabra_{conn}}
{\blabla\det(x-M)\brabra_{conn}} =
\\
= \frac{\p}{\p x} \log \blabla\det(x-M)\brabra_{conn}
\ee
At the same time
the beta-ensemble average of the determinant
$\blabla\det(x-M)\brabra$
is generated (even for both $\epsilon_{1,2}\neq 0$)
by insertion of an additional operator
$V_{1/2b}(x)$ into conformal block (for more precise formulas see
sect.~3.3, (\ref{WyK5})-(\ref{WyKn})):
\be
\blabla\det(x-M)\brabra = \left< V_{1/2b}(x) \ldots \right>
\sim B_5(x|z)
\ee
Therefore, one obtains a much simpler and very transparent SW
representation of the free energy in the $\epsilon_2=0$ limit:\\
i) insert an additional degenerate field $V_{1/2b}(x)$,
i.e. substitute the original $B_4(z)$ by $B_5(x|z)$,\\
ii) consider its asymptotics at small $\epsilon_2$:
\be
B_5(x|z) = \exp \left(-\frac{1}{\epsilon_1\epsilon_2}F(\epsilon_1)
+ \frac{1}{\epsilon_1}S(x;\epsilon_1) + O(\epsilon_2)\right)
\ee
then $dS(x;\epsilon_1)\equiv d_xS(x;\epsilon_1)$ is the generating Seiberg-Witten
differential for
$F(\epsilon_1)\equiv \lim_{\epsilon_2\to 0}\log B_4(z)$
which is the Nekrasov function in the $\epsilon_2\to 0$
by the AGT
relation
\be
a_I = \oint_{A_I} dS(x;\epsilon_1), \nn \\
\frac{\p F(\epsilon_1)}{\p a_I} = b^2\oint_{B_I} dS(x;\epsilon_1)
\ee
since $B_5(x|z)$ satisfies a second-order Schr\"odinger like
differential equation in $x$, the contour integrals can
be considered as Bohr-Sommerfeld periods, describing the monodromy
of the "wave function'' $\psi(x) = B_5(x|z)/B_4(z)$.

The standard dictionary of \cite{intSWd} relates the
supersymmetric Yang-Mills theories with different matter content
with different classical integrable systems,
and $dS_{SW}$ are associated classical short action forms $\vec pd\vec q$,
restricted to spectral curves, while
$F(\epsilon_1)$ arises in this context as the Yang-Yang (YY)-function \cite{NS},
generating the
TBA-like equations of the corresponding quantum integrable system \cite{KT}.
Moreover, in the perturbative limit of gauge theory
the Bohr-Sommerfeld periods of $dS(x;\epsilon_1)$
are given by logarithm of ratios of the
Harish-Chandra functions corresponding to the integrable theory
(i.e. is related to the $S$-matrix).
The $4$-point spherical conformal block captures the family
of $SU(2)$ SYM systems with $N_f\leq 2N_c=4$ fundamental
supermultiplets.
The case of a single adjoint supermultiplet is described by
a parallel theory of the $1$-point toric conformal block
(also with additional insertion of $V_{\BB}(x)$).

We provide more details about this construction
in sect.~4. In particular, an important role is played by
the transparent asymmetry between $\epsilon_1$ and $\epsilon_2$,
both in the Dotsenko-Fateev representation \cite{MMS1}
of the conformal blocks, where only one screening $V_b$
is involved, and in the choice of the degenerate field
$V_{1/2b}(x)$, which is used for insertions.

\section{$B(x|z)$ in CFT}

In this section we describe the standard facts from $2d$ conformal field theory about
the correlators and conformal blocks on sphere and torus with the degenerated field inserted
\cite{CFT,EO} and fix the notation which is used throughout the text.

\subsection{Degenerate primary}

The Verma module $R_\Delta$ generated over the Virasoro highest weight
$V_\Delta = |\Delta\rangle$, $L_nV_\Delta =0$ for $n>0$ and $L_0 V_\Delta = \Delta V_\Delta$,
consists of the linear combinations of the basis vectors $L_{-Y}V_\Delta$.
Here $Y$ denotes arbitrary Young diagram, $Y = \{k_1\geq k_2\geq\ldots\geq k_l>0\}$,
$L_{-Y} \equiv L_{-k_1}\ldots L_{-k_l}$
and $ L_Y = L_{k_l}\ldots  L_{k_1}$.
The Verma module $R_\Delta$ is called degenerate if it contains inside another highest weight
vector $\tilde V = \sum_Y {\tilde C}_YL_{-Y}V_\Delta \neq V_\Delta$, satisfying $ L_n \tilde V = 0$ for $n>0$, (then $\tilde V$ has a vanishing norm).

At the first level $R_\Delta$ is degenerate only if $\Delta = 0$.
If at the second level, $\tilde V = \Big(\xi L_{-1}^2 -  L_{-2}\Big)V_\Delta$
and there are two non-trivial conditions: $ L_1 \tilde V = 0$
and $ L_2 \tilde V = 0$. They imply respectively that
\be
\xi = \frac{3}{2(2\Delta + 1)}
\ee
and
\be
8\Delta + c = 12\xi\Delta
\ee
or, together
\be
\Delta = \frac{5-c \pm \sqrt{(c-1)(c-25)}}{16}
\ee
Parameterizing the central charge and dimension as
\be
\label{cdel}
c = 1 - 6Q^2 = 1 - 6\left(b-\frac{1}{b}\right)^2,  \\
\Delta = \alpha(\alpha - Q) = \alpha\left(\alpha - b + \frac{1}{b}\right)
\ee
we obtain four solutions:
\be
\boxed{
\left\{ \begin{array}{c} \alpha = \frac{1}{2b} \\ \xi = b^2 \end{array}\right.,}\ \ \ \
\left\{ \begin{array}{c} \alpha = -\frac{b}{2}  \\ \xi = \frac{1}{b^2} \end{array}\right.,\ \ \ \
\left\{ \begin{array}{c} \alpha = \frac{3b}{2}-\frac{1}{b}  \\ \xi = b^2 \end{array}\right.,\ \ \ \
\left\{ \begin{array}{c} \alpha = b - \frac{3}{2b}  \\ \xi = \frac{1}{b^2} \end{array}\right.
\ee
In what follows we work with the first of these four solutions  (boxed),
so that the original highest weight primary $V_{\BB}$
of degenerate Verma module satisfies
\be
\tilde V = \Big( b^2 L_{-1}^2 -  L_{-2}\Big)V_{1/2b} = 0
\label{degcon}
\ee
and has dimension
\be
\label{ddg}
\Delta_{1/2b} = -\frac{1}{2} + \frac{3}{4b^2}
\ee
One can impose this constraint on all correlators with insertions of the primary
$V_{\BB}$,
and degeneracy of Verma module implies that this constraint
is a self-consistent requirement.
The conformal Ward identities imply that such correlators satisfy peculiar
differential equations, see sect.~\ref{difeq}.
In the free field realization of CFT this constraint
is imposed almost automatically, see sect.~\ref{chifref}, and this is also
easily seen from the DF/multi-Penner $\beta$-ensemble
representation of the corresponding conformal blocks below in sect.~\ref{3.2.2}.

\subsection{Conformal Ward identities}

The spherical correlators
of primaries satisfy the simple
chain of conformal Ward identities \cite{CFT}:
\be
\left< T(z) \prod_i V_{\alpha_i}(z_i)\right>\
= \left( \sum_i \frac{1}{z-z_i}\p_i + \sum_i \frac{\Delta_i}{(z-z_i)^2}
\right)\left< \prod_i V_{\alpha_i}(z_i)\right>
\label{spheWard}
\ee
and the similar ones for multiple insertions of stress tensor $T(z)$.
In fact, three of the derivatives $\p_i = \p/\p z_i$ in (\ref{spheWard})
can be always eliminated with the help of the projective $SL(2)$-invariance
for the spherical correlators
\be
0 = \ \left< L_{-1}\prod_i V_{\alpha_i}(z_i)\Big)\right>\
= \sum_i \p_i\left< \prod_i V_{\alpha_i}(z_i)\right>\ , \\
0 = \ \left<  L_{0}\prod_i V_{\alpha_i}(z_i)\Big)\right>\
= \sum_i \left(z_i\p_i + \Delta_i\right)
\left< \prod_i V_{\alpha_i}(z_i)\right>\ ,  \\
0 = \ \left<  L_{1}\Big( \prod_i V_{\alpha_i}(z_i)\Big)\right>\
= \sum_i \left(z_i^2\p_i + 2z_i\Delta_i\right)
\left< \prod_i V_{\alpha_i}(z_i)\right>\
\label{SL2inv}
\ee
Equations \rf{spheWard}-\rf{SL2inv} (and similar equations w.r.t. the variables
$\bar z_i$)
for the spherical correlation function holds
for any conformal block $B_I(\{z_i\})$,
with an arbitrary choice of the points $\{ z_i\}$ and intermediate dimensions, which
appear in the channel-decomposition of the correlator
\be
\label{corcb}
\left< \prod_i V_{\alpha_i}(z_i)\right>\ = \sum {\cal C}_{I{\bar J}}B_I(\{z_i\})
{\overline B_J(\{z_i\})}
\ee
i.e. for non-vanishing ${\cal C}_{I{\bar J}}$, where $I$ and ${\bar J}$ are
corresponding holomorphic and anti-holomorphic multi-indices.

In particular, the generic 4-point correlator that solves equations \rf{SL2inv}
can be presented in the form
\be
\left< \prod_{i=1}^4 V_{\alpha_i}(z_i)\right>\ =z_{13}^{-2\Delta_1}
z_{23}^{\Delta_1+\Delta_4-\Delta_2-\Delta_3}
z_{34}^{\Delta_1+\Delta_2-\Delta_3-\Delta_4}
z_{24}^{\Delta_3-\Delta_1-\Delta_2-\Delta_4}\times (\bar z \hbox{ part})
\times G(x,\bar x)
\ee
where $z_{ij}\equiv z_i-z_j$ and
$G$ is the function of only the double ratios $x={z_{12}z_{34}\over
z_{13}z_{24}}$ and similarly for
$\bar x$. This allows one to choose the fields located at $z_1=0$, $z_2=x$, $z_3=1$ and $z_4=\infty$,
the 4-point conformal block in formula \rf{corcb} acquires the form
\be
\label{B4pt}
B_\Delta(x) \equiv B_\Delta^{(12;34)}(x) = x^{\Delta-\Delta_1-\Delta_2}
\sum_{n>0} B_{\Delta,n}x^n =
\\
= x^{\Delta-\Delta_1-\Delta_2}\left(1 + {(\Delta+\Delta_1 -\Delta_2   )(\Delta+\Delta_3-\Delta_4)
\over 2\Delta}x + \ldots\right)
\ee

\subsection{Equation for the conformal block
\label{difeq}}

For our purposes in this paper we distinguish one of the primaries,
$V_{\alpha_0}(x)$ at some point $z_0=x$, with dimension $\Delta_0 = \Delta(\alpha_0)$,
which later will be made degenerate at the second level.
Integrating \rf{spheWard} over $z$ with the weight $(z-x)^{-1}$, one obtains
\be
\label{l2}
\left<  L_{-2} V_{\alpha_0}(x) \prod_i V_{\alpha_i}(z_i)\right>\ =
\left(
\sum_i \frac{1}{x-z_i}\p_i + \sum_i \frac{\Delta_i}{(x-z_i)^2}
\right)\left< V_{\alpha_0}(x) \prod_i V_{\alpha_i}(z_i)\right>\
\ee
and, similarly,
\be
\label{l1}
\left<  L_{-1}^2 V_{\alpha_0}(x) \prod_i V_{\alpha_i}(z_i)\right>\ =
\p_x^2 \left< V_{\alpha_0}(x) \prod_i V_{\alpha_i}(z_i)\right>\
\ee
Choosing $\alpha_0={1\over 2b}$
and making use of (\ref{degcon}), one gets that
\be
\left( b^2\p_x^2 -
\sum_i \frac{1}{x-z_i}\p_i - \sum_i \frac{\Delta_i}{(x-z_i)^2}
\right)\left< V_{\BB}(x) \prod_i V_{\alpha_i}(z_i)\right>\ =0
\label{degdifeq}
\ee
Now we apply this equation to the conformal block and realize that it fixes
a specific intermediate dimensions in the conformal block.

\subsubsection{Four-point conformal block with the degenerate field
\label{ss:cft4}}

Equations \rf{SL2inv} are enough to reduce (\ref{degdifeq})
to a {\it single}-variable differential equation
in the case of only {\it three} variables $z_{1,2,3}$:
the three equations (\ref{SL2inv}) allow to express all the three
derivatives $\p_i$.
Substituting these expressions back into (\ref{degdifeq}),
one obtains \cite{CFT}:
\be
\left\{ b^2\p_x^2 + \sum_{i=1}^3{1\over x-z_i}\p_x +
\frac{3x-z_1-z_2-z_3}{(x-z_1)(x-z_2)(x-z_3)}\Delta_{1/2b}
+ \frac{(z_1-z_2)(z_3-z_1)\Delta_1}{(x-z_1)^2(x-z_2)(x-z_3)}
 \right.  \\ \left.
+ \frac{(z_1-z_2)(z_2-z_3)\Delta_2}{(x-z_1)(x-z_2)^2(x-z_3)} +
\frac{(z_2-z_3)(z_3-z_1)\Delta_3}{(x-z_1)(x-z_2)(x-z_3)^2}
\right\}B_4(x|z_1,z_2,z_3) = 0 \ee If $z_{1,2,3}$ are placed at
$0,1,\infty$, then this equation simplifies to \be \label{hg1}
\left\{ b^2x(x-1)\p_x^2 + (2x-1)\p_x + \Delta_{1/2b} +
\frac{\Delta_1}{x} - \frac{\Delta_2}{x-1} - \Delta_3\right\}
B_4(x|0,1,\infty) =0 \ee Conjugation with a factor
$x^\alpha(1-x)^\beta$ with specially adjusted $\alpha$ and $\beta$
converts this into an ordinary hypergeometric equation with the
solution \be \label{hg2} B_4(x|0,1,\infty) =
x^{\alpha_1/b}(1-x)^{\alpha_2/b}F(A,B;C;x)
\\
A = {1\over 2b^2}+{\alpha_1\over b}+{\alpha_2\over b}-{\alpha_3\over b}
\\
B = {1\over b}\sum_{i=1}^3\alpha_i + 2\Delta_{1/2b},\ \ \ C={1\over b^2} + {2\alpha_1\over b}
\ee
Equations \rf{hg1}, \rf{hg2} are consistent with generic formula \rf{B4pt} {\it only}
if the dimensions $\Delta_1$ and $\Delta$
are related by the fusion rule\footnote{
We specially stress this point here: in the original normalizations of \cite{CFT}, when the
structure constants are not absorbed into the definition of conformal block, generic formula
\rf{B4pt} {\it by no means simplifies}, or, e.g. as is stated in Appendix of \cite{CDV} gives rise
to vanishing result. Expression \rf{B4pt} simplifies only upon conditions \rf{fusion} and all
other conformal blocks, though being non-vanishing themselves, do not give contributions to the
correlator due to vanishing of the structure constants, see \rf{ccb4dg}.}
\be
\label{fusion}
\alpha = \alpha_1 \pm {1\over 2b}
\\
\Delta_1 = \alpha_1\left(\alpha_1 - b + \frac{1}{b}\right),\ \ \
\Delta = \Delta_\alpha = \alpha\left(\alpha - b + \frac{1}{b}\right)
\ee
where two choices of the sign correspond to the two linearly independent solutions of
\rf{hg1} and in the case of the sign ``minus'' in (\ref{fusion}) one has to choose
in (\ref{hg2}) instead of $F(A,B;C;x)$ the other
solution to the hypergeometric equation $x^{1-C}F(A-C+1,B-C+1;2-C;x)$.

One can easily check directly that the conformal block from the r.h.s. of
\rf{ccb4dg}
\be
\label{B4ptdg}
B_{\Delta_\alpha}^{(1,{1/2b};34)}(x) = x^{\Delta_\alpha-\Delta_1-\Delta_{1/2b}}
\left(1 + {(\Delta_\alpha+\Delta_{1/2b}-\Delta_1)(\Delta_\alpha+\Delta_3-\Delta_4)
\over 2\Delta_\alpha}x + \ldots\right) =
\\
\stackreb{\rf{fusion}}{=}\ B_4(x|0,1,\infty)
\ee
which solves \rf{hg1}. Formula \rf{corcb} now acquires the form
\be
\label{ccb4dg}
\left< V_1(0)V_{1/2b}(x)V_3(1)V_4(\infty)\right> =
\sum_\Delta C_{1,{1/2b}}^\Delta C_{34}^\Delta
\left|B_\Delta^{(1,{1/2b};34)}(x)\right|^2 =
\\
= \sum_{\alpha = \alpha_1 \pm {1\over 2b}} C_{1,{1/2b}}^{\Delta_\alpha} C_{34}^{\Delta_\alpha}
\left|B_{\Delta_\alpha}^{(1,{1/2b};34)}(x)\right|^2
\ee
since {\em only} for the choice \rf{fusion} the structure constant $C_{1,{1/2b}}^{\Delta_\alpha}$
is nonvanishing \cite{CFT}. Here we obtained this fact indirectly by solving the
equation for the correlator. We shall derive this fact straightforwardly
using the $\beta$-ensemble representation for the conformal blocks in the next section.

\subsubsection{Five-point conformal block with the degenerate field\label{ss:cft5}}

When there are {\it four} variables $z_{1,2,3,4}$, then one can use
(\ref{SL2inv}) to eliminate three out of four derivatives $\p_i$:
\be
\left\{ b^2\p_x^2 + \frac{3x^2-2x(z_1+z_2+z_3)+z_1z_2+z_2z_3+z_3z_1}
{(x-z_1)(x-z_2)(x-z_3)}\,\p_x
+ \frac{(z_1-z_4)(z_2-z_4)(z_3-z_4)}{(x-z_1)(x-z_2)(x-z_3)(x-z_4)}\p_4 +
\phantom{5^{5^{5^{5^{5^{5^{5^{5^5}}}}}}}}
\!\!\!\!\!\!\!\!\!\!\!\!\!\!\!\!\!\!\!\!\!\!
\right.  \\  \\ \left.
+ \frac{(z_1-z_2)(z_2-z_3)(z_3-z_1)}{(x-z_1)(x-z_2)(x-z_3)}
\left(\frac{\Delta_1}{(x-z_1)(z_2-z_3)} + \frac{\Delta_2}{(x-z_2)(z_3-z_1)} +
\frac{\Delta_3}{(x-z_3)(z_1-z_2)}\right)
- \right.  \\  \\ \left.
- \frac{\Big(3z_4^2-2z_4(z_1+z_2+z_3)+z_1z_2+z_2z_3+z_3z_1\Big)x -
\Big(2z_4^3 - (z_1+z_2+z_3)z_4^2 + z_1z_2z_3\Big)
}{(x-z_1)(x-z_2)(x-z_3)(x-z_4)^2}\Delta_4
+ \right.  \\  \\ \left.
+ \frac{3x-z_1-z_2-z_3}{(x-z_1)(x-z_2)(x-z_3)}\Delta_{1/2b}
\right\}
B_5(x|z_1,z_2,z_3,z_4) = 0
\ee
If $z_{1,2,3}$ are placed at $0,1,\infty$, this equation for $B(x|0,1,\infty,q)
\equiv B(x|q)$ simplifies to

\fr{
\label{29}
\left\{ b^2x(x-1)\p_x^2 + (2x-1)\p_x - \frac{q(q-1)}{x-q}\p_q
+ \Delta_{1/2b}
+ \frac{\Delta_1}{x} - \frac{\Delta_2}{x-1} - \Delta_3
+ \frac{q^2-(2q-1)x}{(x-q)^2}\Delta_4
\right\} \\
B_5(x|q) =0}
$x$ and $x-1$ in denominators can be again eliminated by conjugation.
Resulting equation can be represented as the one on an elliptic curve
(torus) with coordinate $x-q$ and ramification point $q^{-1}$ \cite{ZFLON}.
The double pole $(x-q)^2$ then becomes a Weierstrass function.

\subsubsection{Toric block with one $z$-variable}

Instead of (\ref{spheWard}) and (\ref{SL2inv})
a toric correlator satisfies a pair of equations: the conformal Ward identity
\cite{EO} (we normalize the correlators so that the toric partition function is
$Z(\tau,{\bar\tau})=\left< 1\right>$)
\be
\left< T(z) \prod_i V_{\alpha_i}(z_i)\right> =
2\pi i{\d\over\d\tau}\left< \prod_i V_{\alpha_i}(z_i)\right>+
\\
+
\sum_i \Big((\zeta_*(z-z_i|\tau)+2\eta_1z)\p_i + \Delta_i\wp_*(z-z_i|\tau)\Big)
\left< \prod_i V_{\alpha_i}(z_i)\right>
\label{tward}
\ee
with
\be
\zeta_*(z|\tau) \equiv \p_z\log\theta_*(z|\tau)=\zeta(z|\tau)-2\eta_1 z,
\ \ \ \ \
\wp_*(z|\tau) \equiv -\p_z\zeta_*(z|\tau) = \wp(z|\tau)+2\eta_1\\
\eta_1=\zeta(\half|\tau)=
-2\pi i\d_\tau\log\eta\left(e^{i\pi\tau}\right),\ \ \ \ \ \ \
\eta(\tau) = e^{i\pi\tau/12}\prod_{n>0}\left(1-e^{2in\pi\tau}\right)
\ee
and torus counterpart of (\ref{SL2inv}) looks as
\be
\label{torl1}
0 = \left<   L_{-1}\Big( \prod_i V_{\alpha_i}(z_i)\Big)\right>
= \sum_i \p_i\left< \prod_i V_{\alpha_i}(z_i)\right>
\ee
Note that \rf{torl1} ensures correctness the double-periodicity in $z$
of the equation \rf{tward}, while for the periodicity in $\{ z_i\}$-variables
the presence of $\tau$-derivative is extremely important.

As a corollary of (\ref{degcon}) and \rf{tward},
one obtains a torus counterpart of eq.(\ref{degdifeq}):
the correlator with the degenerate field insertion now satisfies
\be
\left(-2\pi i{\d\over\d\tau} + b^2\p_x^2  -
\sum_j \left(\zeta_*(x-z_j|\tau)\p_j + \Delta_j\wp_*(x-z_j)\right)
\right)\left< V_{\BB}(x) \prod_i V_{\alpha_i}(z_i)\right> =
\\
=2\eta_1\Delta_{1/2b}\left< V_{\BB}(x) \prod_i V_{\alpha_i}(z_i)\right>
\label{degdifeqtor}
\ee
In the particular case of a single $z$-variable (to be put at $z=0$) we get:
\be
\left(-2\pi i{\d\over\d\tau} + b^2\p_x^2 +
 \zeta_*(x|\tau)\p_x - \Delta_\alpha\wp_*(x)
\right)
\left< V_{\BB}(x)  V_{\alpha}(0)\right> =
 2\eta_1\Delta_{1/2b}\left< V_{\BB}(x)  V_{\alpha}(0)\right>
\label{degdifeqtor1}
\ee
or, after multiplication by $\eta^A\theta_\ast(x)^{-1/2b^2}$ with
${A\over 2} =\Delta_\alpha+{1\over b^2}-1$, it turns into
(cf. e.g. with \cite{FLit1})
\be\label{34}
\boxed{
\left(2\pi i{\d\over\d\tau} - b^2\p_x^2
 +\left(\Delta_\alpha+{1\over 4b^2}-{1\over 2}\right)\wp(x)\right)\cdot
\left(\eta^{-A}\theta_\ast(x)^{1/2b^2}\left< V_{\BB}(x)  V_{\alpha}(0)\right>\right)=0}
\ee
and the same equation is satisfied by any toric 2-point conformal block, arising
in the decomposition of the correlator
\be
\label{tordec}
\left< V_{\BB}(x)  V_{\alpha}(0)\right> = \sum_{\Delta,\pm} C_{\Delta_\alpha\Delta}^{\Delta_\pm}\left|B_{\Delta,\Delta_\alpha}^\pm(x|\tau)\right|^2
\ee
with
\be
\Delta_\pm = \Delta_{\alpha_\pm},\ \ \ \
\alpha_\pm = \alpha\pm{1\over 2b}
\ee

\subsubsection{The "non-conformal" limit
\label{nonconf}}

When the four external dimensions $\Delta(\alpha_i)$ become large (while
the intermediate dimension $\Delta$ is kept finite), one can make the
double ratio $q = \frac{(z_2-z_1)(z_4-z_3)}{(z_3-z_1)(z_4-z_2)}$
small, so that the dimensional transmutation takes place, and a new
finite parameter $\Lambda^4 =
q\sqrt{\Delta_1\Delta_2\Delta_3\Delta_4}$ emerges instead of $q$ and
four $\Delta_i$. This limit corresponds to the pure ${\cal N}=2$ $SU(2)$
SYM theory and thus is referred to as "non-conformal limit" in AGT
literature, see \cite{nonconf}. On the CFT side, this limit is
associated with a peculiar coherent state
\be
|\Delta,\Lambda\rangle\ = \sum_Y
\Lambda^{2|Y|}Q^{-1}_\Delta\Big([1^{|Y|}],Y\Big)L_{-Y}|\Delta\rangle
\ee
so that the 4-point conformal block turns into
\be
B_\Delta^{12;34}(q) \rightarrow\
\left<\Delta,\Lambda|\Delta,\Lambda\right> = \sum_{n\geq 0}
\Lambda^{4n}Q^{-1}([1^n],[1^n])
\ee
where the sum goes over the
single-row Young diagrams $Y = [1^n]$, and $Q(Y,Y') = \left<\Delta|
L_Y L_{-Y'}|\Delta\right>$ is the block-diagonal Shapovalov form for
the Virasoro algebra. The same result can be of course obtained from
a similar limit of the 1-point toric conformal block
$B_{\Delta,\Delta_\alpha}(\tau)$, which corresponds on the SYM side
to obtaining the pure gauge theory from the infinite-mass limit $e^{\pi
i\tau}\Delta_\alpha=\Lambda^2$ (being fixed when
$\Delta_\alpha\to\infty$ and $\tau\to +i\infty$) of the ${\cal
N}=2^*$ theory with adjoint supermultiplet:
\be\label{32}
B_{\Delta,\Delta_\alpha}(\tau) \rightarrow\
\left<\Delta,\Lambda|\Delta,\Lambda\right> = \sum_n
\Lambda^{4n}Q^{-1}([1^n],[1^n])
\ee

In this paper we are interested in the conformal block with additional
insertion of the degenerate primary $V_\BB(x)$. There are three possibilities to obtain
the equation for this
conformal block. First of all, one can obtain the equation directly by insertion
of the degenerate primary into the matrix element (\ref{32}):
\be
{\cal B}_5(x|z_1,z_2,z_3,z_4) \rightarrow\
\left<\Delta,\Lambda|V_\BB(x)|\Delta,\Lambda\right>
\ee
This was done in \cite{5dJ}. The two other possibility are those which we discussed above:
one can take the limit of infinite
masses in equation (\ref{29}) for the 5-point conformal block $B_5$, or
consider a similar limit for the toric
conformal block \rf{tordec},
\be
B_{\Delta,\Delta_\alpha}^\pm(x|\tau) \rightarrow\
\left<\Delta,\Lambda|V_\BB(x)|\Delta,\Lambda\right>
\ee
All three methods definitely lead to the same equation.
For instance, in the latter case eqn.(\ref{34}) is substituted by its periodic Toda-chain
(sine-Gordon) analogue.
Indeed, in the peculiar Inozemtsev limit \cite{Ino} the Weierstrass function
$\wp(x|\tau)$ turns into a hyperbolic cosine. To see this, rewrite first the
Weierstrass function as an expansion in inverse sines:
\be
\wp(x) = \sum_{m,n\in\ \mathbb{Z}}\frac{1}{(x+m+n\tau)^2} - C(\tau) = \sum_{n\in\ \mathbb{Z}}
\frac{\pi^2}{\sin^2\pi(x+n\tau)} - C(\tau)
\ee
(where the factor $\pi$ emerges in the argument of $\sin$ due to periodicity under
$x\rightarrow x+i\pi$, while the factor $\pi$ in the numerator is present since
$\frac{\pi}{\sin\pi x} \sim \frac{1}{x}$), and
\be
C(\tau)={1\over 3}+2\sum_{n\ge 1}
\frac{\pi^2}{\sin^2\pi(n\tau)}
\ee
Next, put $x = i\xi -\tau/2$. In the Inozemtsev limit there are two terms, surviving from this
sum in the leading order in $e^{2\pi i \tau}$-expansion:
\be
\frac{\pi^2}{\sin^2 \pi x} \longrightarrow -4\pi^2
e^{i\pi\tau}e^{-2\pi \xi}
\ee
and
\be
\frac{\pi^2}{\sin^2 \pi (x+\tau)} \longrightarrow -4\pi^2
e^{i\pi\tau}e^{+2\pi \xi}
\ee
Thus
\be
\wp(x) \rightarrow -8\pi^2 e^{\pi i\tau} \cosh{2\pi\xi}
\ee
Of course, $\p_x = -i\p_\xi$ and
the Calogero-Shr\"odinger
equation (\ref{34}) finally turns into (under the rescaling $2\pi\xi\to\xi$)

\be\label{nonconf2}
\boxed{
\left( b^2\p_\xi^2 - 2\Lambda^2\cosh \xi +{1\over 4}\frac{\p}{\p\log\Lambda}
\right)
\left<\Delta,\Lambda|V_\BB(\xi)|\Delta,\Lambda\right> = 0}
\ee
\smallskip

\noindent
This formula coincides with \cite[(A.13)]{5dJ} up to some trivial rescalings of the
conformal block.

\section{$B(x|z)$ in free field/$\beta$-ensemble realizations}

\subsection{Free fields \cite{FreeF,Bos}
\label{chifref}}

The chiral free field propagator is given by
\be
\label{ff}
\left<\phi(z) \phi(0)\right> = -2\log z
\ee
For the exponential primary fields
\be
V_\alpha =\ : e^{i\alpha\phi}:
\ee
one can write
\be
\prod_j V_{\alpha_j}(z_j) = \prod_j :e^{i\alpha_j(z_j)}:\ =
\prod_{i<j} (z_i-z_j)^{2\alpha_i\alpha_j}
:e^{\sum_j\alpha_j \phi(z_j)}:\ =
\\
= \prod_{i<j} (z_i-z_j)^{2\alpha_i\alpha_j}:\prod_j V_{\alpha_j}(z_j) :
\label{confcor}
\ee
The holomorphic stress-tensor
\be
T = -\frac{1}{4}(\p\phi)^2 + \frac{iQ}{2}\p^2\phi
\ee
obviously satisfies
\be
T(z)T(0) = \frac{c}{2z^4} + \frac{2}{z^2}T(0) + \frac{1}{z}\p T(0) + O(z)
\ee
and
\be
T(z)V_\alpha(0) = \sum_k \frac{1}{z^{k+2}}  L_kV_\alpha(0)
= \frac{\Delta_\alpha}{z^2}V_\alpha(0) + \frac{1}{z}\p V_\alpha(0)
+
\\
+ :\!\!\left(-\frac{1}{4}(\p\phi)^2 + i\left(\alpha
+\frac{Q}{2}\right)\p^2\phi\right)
V_\alpha(0): + O(z)
\label{tv}
\ee
with the central charge and dimension exactly given by \rf{cdel}.
The screening currents with unit dimension are $V_b$ and $V_{-1/b}$,
since $\Delta_b = \Delta_{-1/b} = 1$.

The null-vector condition implies that
\be
\label{nvff}
\Big(b^2 L_{-1}^2 -  L_{-2}\Big)V_\alpha =
b^2\p^2 V_\alpha -
:\!\!\left(-\frac{1}{4}(\phi)^2 + i\left(\alpha+\frac{Q}{2}\right)\p^2\phi\right)
V_\alpha: \ =  \\ =
\ :\left(
\left(\alpha^2b^2-\frac{1}{4}\right)(\p\phi)^2 +
i\left(\alpha b^2-\alpha - \frac{Q}{2}\right)\p^2\phi\right)V_\alpha:
\ee
and the r.h.s.
vanishes for $\alpha = \frac{1}{2b}$ (and $Q=b-\frac{1}{b}$). In what follows we shall
omit the normal-ordering signs for the free-field operators, when their presence
is obvious.

\subsection{$B(x|z)$ in the $\beta$-ensemble representation
\label{beten}}

\subsubsection{Conformal block in the free field representation}

In the free field realization the arbitrary generic conformal block on sphere
is given by
\be
B_I(\{ z_i\}) =
\la \prod_i
e^{i\alpha_i\phi(z_i)} \prod_{\gamma_I}
\left(\int e^{ib\phi(u)} du\right)^{N_{\gamma_I}}\ra_{free}
\ee
where angular brackets imply the correlator in the theory of
$2d$ chiral field \rf{ff}
according to the rule (\ref{confcor}).
The numbers of screening insertions $N_{\gamma_I}$ and the choice of the integration contours
$\gamma_I$ themselves depend on particular choice $I$ of the conformal block.

For example, in the case of 4-point conformal block one can write
\be
B_\Delta^{(12;34)}(q) =
\la
e^{i\alpha_1\phi(0)}e^{i\alpha_2\phi(q)}
e^{i\alpha_3\phi(1)}e^{i\alpha_4\phi(\infty)}
\left(\int_0^q e^{ib\phi(u)}du\right)^{N_x}
\left(\int_0^1 e^{ib\phi(v)}dv\right)^{N_1} \ra_{free} =
\\
= q^{2\alpha_1\alpha_2}(1-q)^{2\alpha_2\alpha_3}
\int \prod_{a<a'} (U_a-U_{a'})^{2b^2}
\prod_a U_a^{2\alpha_1b}(1-U_a)^{2\alpha_3b}
(q-U_a)^{2\alpha_2b} dU_a
\label{Bff4}
\ee
where $\{ U_a\} = \{ \{u\},\{v\}\}$, $a=1,\ldots,N_x+N_1$, with
\be
N_x = {1\over b}\left(\alpha-\alpha_1-\alpha_2\right)
\\
N_1 = {1\over b}\left(Q-\alpha-\alpha_3-\alpha_4\right)
\ee
being the number of contours stretched between $u=0,q$ and $v=0,1$ correspondingly.

\subsubsection{Four-point conformal block in the $\beta$-ensemble representation
\label{3.2.2}}

The purpose of this section is to show that the differential equation
(\ref{degdifeq}) survives for the conformal block only provided that no
screening contour terminates at position of the degenerate operator.

If one of the fields is degenerate at the second level, say $V_\BB(x)$, formula
\rf{Bff4} gives the solution to the second-order null-vector equation \rf{hg1}
only for $N_x=0$, i.e.
\be
B_4(x|0,1,\infty) =
\la
e^{{i\over 2b}\phi(x)}e^{i\alpha_1\phi(0)}
e^{i\alpha_2\phi(1)}e^{i\alpha_3\phi(\infty)}
\left(\int_0^1 e^{ib\phi(v)}dv\right)^{N_1}
\ra_{free} =
\\
= x^{\alpha_1/b}(1-x)^{\alpha_2/b}
\int \prod_{a<a'} (v_a-v_{a'})^{2b^2}
\prod_a v_a^{2\alpha_1b}(1-v_a)^{2\alpha_2b}
(x-v_a) dv_a
\\
N_1 = {1\over b}\left(Q-\alpha-\alpha_3-\alpha_4\right)
\label{Bff4dg}
\ee
Equation \rf{hg1} follow for the r.h.s. of \rf{Bff4dg} automatically when
applying \rf{nvff} for the field $V_{1/2b}(x)$. Consider the free-field correlator
with one degenerate field and some number of screenings (with unspecified yet contours)
inserted, then
\be
\la
\left\{(b^2 L_{-1}^2 -  L_{-2})e^{{i\over 2b}\phi(x)}\right\}e^{i\alpha_1\phi(0)}
e^{i\alpha_2\phi(1)}e^{i\alpha_3\phi(\infty)}
\prod_{I}
\left(\int_{\gamma_I} e^{ib\phi(u)} du\right)^{N_{\gamma_I}}
\ra_{free} = 0
\ee
is obviously true, due to equation \rf{nvff} at $\alpha={1\over 2b}$. As for an arbitrary
conformal theory above, one can write
\be
b^2\p_x^2\la
 e^{{i\over 2b}\phi(x)}e^{i\alpha_1\phi(0)}
e^{i\alpha_2\phi(1)}e^{i\alpha_3\phi(\infty)}
\prod_{I}
\left(\int_{\gamma_I} e^{ib\phi(u)} du\right)^{N_{\gamma_I}}
\ra_{free} =
\\
=b^2\la
 L_{-1}^2e^{{i\over 2b}\phi(x)}e^{i\alpha_1\phi(0)}
e^{i\alpha_2\phi(1)}e^{i\alpha_3\phi(\infty)}
\prod_{I}
\left(\int_{\gamma_I} e^{ib\phi(u)} du\right)^{N_{\gamma_I}}
\ra_{free} =
\\
=\la
L_{-2}e^{{i\over 2b}\phi(x)}e^{i\alpha_1\phi(0)}
e^{i\alpha_2\phi(1)}e^{i\alpha_3\phi(\infty)}
\prod_{I}
\left(\int_{\gamma_I} e^{ib\phi(u)} du\right)^{N_{\gamma_I}}
\ra_{free} =
\\
= \oint_x {dz\over z-x}\la
\left(T(z)e^{{i\over 2b}\phi(x)}\right)e^{i\alpha_1\phi(0)}
e^{i\alpha_2\phi(1)}e^{i\alpha_3\phi(\infty)}
\prod_{I}
\left(\int_{\gamma_I} e^{ib\phi(u)} du\right)^{N_{\gamma_I}}
\ra_{free} =
\\
= - \sum_{w=0,1,\infty}
\oint_x {dz\over z-x}\la
e^{{i\over 2b}\phi(x)}\left(T(z)e^{i\alpha_1\phi(0)}
e^{i\alpha_2\phi(1)}e^{i\alpha_3\phi(\infty)}
\prod_{I}
\left(\int_{\gamma_I} e^{ib\phi(u)} du\right)^{N_{\gamma_I}}\right)
\ra_{free}
\label{tscor}
\ee
The r.h.s. of this equation obviously results in the expression as in r.h.s. of \rf{l2},
giving rise further to \rf{hg1}, \rf{hg2} exactly as in sect.~\ref{ss:cft4},
but now directly for the particular conformal block, written in the form of the
free-field correlator \rf{Bff4dg}. One has to use here only commutativity of
the stress-energy tensor with the screening operator $[T(z),\int V_b(u)du]=0$,
following from the fact that the singular part of the corresponding OPE \rf{tv}
\be
T(z)V_b(u) = {\p\over\p u}\left({V_b(u)\over z-u}\right) + \ldots
\label{tscom}
\ee
is total derivative.

This argument should be applied however with an extra care in the case of
non-closed contours (like in \rf{Bff4} and \rf{Bff4dg}). Usually  (see e.g. \cite{confmamo})
the desired result is achieved by the analytic continuation of the result of free-field
computation from the values of parameters, ensuring automatic vanishing of this result
at the end-point of the contour. However, this is not always possible, and we shall see
immediately, that in our case the argument is applicable only for $N_x=0$, i.e. when
there is no integration of the screening current with the end-point at $u=x$.

Indeed, due to \rf{tscom}, the integrand in the correlator \rf{tscor} contains the term of the form
\be
{\p\over\p u}\cdot{1\over x-u}\la
e^{{i\over 2b}\phi(x)}e^{i\alpha_1\phi(0)}
e^{i\alpha_2\phi(1)}e^{i\alpha_3\phi(\infty)}
{e^{ib\phi(u)}} \ra_{free}\ \stackreb{\rf{Bff4dg}}{ =}
\\
= {\p\over\p u}\left({1\over x-u}\cdot u^{2\alpha_1b}(1-u)^{2\alpha_2b}
(x-u)\right) = {\p\over\p u}\left(u^{2\alpha_1b}(1-u)^{2\alpha_2b}\right)
\label{dscrp}
\ee
i.e. the extra pole from \rf{tscom} exactly cancels the zero at $x=u$, coming
from the contraction of degenerate field with the screening current
\be
\label{main}
V_{1/2b}(x)\cdot V_b(u)= e^{{i\over 2b}\phi(x)}\cdot {e^{ib\phi(u)}} \stackreb{\rf{ff}}{\sim} (x-u)
\ee
The integral of \rf{dscrp} along the contour between $u=0$ and $u=x$ is obviously nonvanishing,
contrary to the integral between $u=0$ and $u=1$, which can be treated as vanishing at least in
the sense of analytic continuation. We therefor conclude, that \rf{Bff4} satisfies the second-order
differential equation in $x$-variable, if the field at the point $x$ is degenerate on second level
{\em and} $N_x=0$, i.e. exactly for \rf{Bff4dg}.

\subsubsection{$n$-point conformal block in the $\beta$-ensemble representation}

For the 5-point conformal block with $q\ll x\ll 1$,
$q = \frac{(z_2-z_1)(z_3-z_4)}{(z_3-z_1)(z_2-z_4)}$, see Fig.\ref{conf41},
one has \cite{MMS1,MMS}
\be
B_5(x|q) = \ \la
:e^{i\alpha_1\phi(0)}: \
:e^{i\alpha_2\phi(q)}: \
:e^{\frac{i}{2b}\phi(x)}:\
:e^{i\alpha_3\phi(1)}: \
:e^{i\alpha_4\phi(\infty)}: \right.  \\ \left.
\left(\int_0^q :e^{ib\phi(u)}:\ du\right)^{N_q}
\left(\int_0^x :e^{ib\phi(v)}:\ dv\right)^{N_x}
\left(\int_0^1 :e^{ib\phi(w)}:\ dw\right)^{N_1}
\ra_{free}
\label{Bthrprim}
\ee
where
\be
\alpha = \alpha_1+\alpha_2 + bN_q,  \\
{\tilde\alpha}=\alpha + \frac{1}{2b} + bN_x,  \\
\alpha_4 = b-\frac{3}{2b} - \alpha_1-\alpha_2-\alpha_3 -
b(N_q+N_x+N_1)
\ee
Here the angular brackets imply just the free field computation (\ref{confcor}), i.e.
\be
B_5(x|q) \sim q^{2\alpha_1\alpha_2}(1-q)^{2\alpha_2\alpha_3}
x^{\alpha_1/b}(1-x)^{\alpha_3/b}(q-x)^{\alpha_2/b}\cdot \\ \cdot
\int \underline{\prod_i (x-U_i)}  \prod_{i<j} (U_i-U_j)^{2b^2}
\prod_i U_i^{2\alpha_1b}(1-U_i)^{2\alpha_3b}
(q-U_i)^{2\alpha_2b} dU_i
=\\=
e^{{1\over 2b^2}W(x)}B_4(q)
{\int \underline{\prod_i (x-U_i)}  \prod_{i<j} (U_i-U_j)^{2b^2}
\prod_i
%U_i^{2\alpha_1b}(1-U_i)^{2\alpha_3b}
%(q-U_i)^{2\alpha_2b}
e^{W(U_i)}dU_i
\over
\int \prod_{i<j} (U_i-U_j)^{2b^2}
\prod_i
%U_i^{2\alpha_1b}(1-U_i)^{2\alpha_3b}
%(q-U_i)^{2\alpha_2b}
e^{W(U_i)}dU_i}=\\=
\exp\left({1\over 2b^2}\int^x W'({\tilde x})d{\tilde x}\right)B_4(q)
\blabla \underline{"\det"(x-M)} \brabra
\label{Bdet}
\ee
where we used (\ref{Bff4}), and
\be
\label{betapot}
W(x)=2b\sum_{i=1}^4 \alpha_i\log (x-z_i)
\\
(i=1,2,3,4,\ \  z_i=\{0,q,1,\infty\})
\ee
is the logarithmic potential of the
beta-ensemble with $\beta=b^2$, corresponding to (\ref{Bff4}).
As usual in, we symbolically denote the r.h.s. as a "matrix-model"
average, as if $M$ was a "matrix" with eigenvalues $U_i$:
$M = {\rm diag}(U_i) = {\rm diag}(\{u\},\{v\},\{w\})$,
determinant $"\det"(x-M) \equiv \prod_i(x-U_i)$ and integration
measure $dM \equiv \prod_{i<j} (U_i-U_j)^{2\beta}\prod_i dU_i$.
Double angular brackets denote the beta-ensemble average
with specific integration contours, different for three
different constituents of the "eigenvalue set" $\{U_a\}$,
as in eq.(\ref{Bthrprim}).

As in the case of 4-point function \rf{Bff4dg} it is easy to check
that this multiple integral indeed
satisfies (\ref{degdifeq}), but only for $N_x=0$, which also implies the same
fusion rule as in the case of the four-point conformal block.

Similarly, the $n$-point conformal block with $n-4$ degenerated operators
is given for the comb-like diagram \cite{MMS1,MMS}
by the free field average, $q\ll x_1\ll\ldots\ll x_{n-4}\ll 1$
\be
B_n(x_a|q) = \ \la
:e^{i\alpha_1\phi(0)}: \
:e^{i\alpha_2\phi(q)}: \
\prod_a :e^{\frac{i}{2b}\phi(x_a)}:\
:e^{i\alpha_3\phi(1)}: \
:e^{i\alpha_4\phi(\infty)}: \right.  \\ \left.
\left(\int_0^q :e^{ib\phi(u)}:\ du\right)^{N_q}
\prod_a\left(\int_0^{x_a} :e^{ib\phi(v_a)}:\ dv_a\right)^{N_{x_a}}
\left(\int_0^1 :e^{ib\phi(w)}:\ dw\right)^{N_1}
\ra_{free}
\label{Bthrprimn}
\ee
with
\be\label{consl}
\alpha = \alpha_1+\alpha_2 + bN_q,  \\
{\alpha}^{(a)}=\alpha^{(a-1)} + \frac{1}{2b} + bN_{x_a},  \\
\alpha_4 = b-\frac{n-2}{2b} - \alpha_1-\alpha_2-\alpha_3 -
b(N_q+\sum_aN_{x_a}+N_1)
\ee
where $\alpha^{(a)}$ refers to the intermediate channels.

The eigenvalue model average now looks like
\be\label{Bdetn}
B_n(x_a|q)\sim
q^{2\alpha_1\alpha_2}(1-q)^{2\alpha_2\alpha_3}
\prod_a x_a^{\alpha_1/b}(1-x_a)^{\alpha_3/b}(q-x_a)^{\alpha_2/b}
\prod_{a<b}(x_a-x_b)^{{1\over b^2}}\cdot \\ \cdot
\int \underline{\prod_a\prod_i (x_a-U_i)}  \prod_{i<j} (U_i-U_j)^{2b^2}
\prod_i U_i^{2\alpha_1b}(1-U_i)^{2\alpha_3b}
(q-U_i)^{2\alpha_2b} dU_i=\\=
\exp\left(\sum_a{1\over 2b^2}\int^{x_a} W'({\tilde x})d{\tilde x}\right)B_4(q)
\blabla \underline{\prod_a "\det"(x_a-M)} \brabra
\ee
Again this multiple integral indeed
satisfies (\ref{degdifeq}), but only for all $N_{x_a}=0$, which also implies the same
fusion rule as in the case of the four-point conformal block.

\subsection{Resolvent expansion of the conformal block}

Applying general identity
\be
\log \blabla e^L \brabra\ = \log \left(1 + \lala\ L\ \rara
+ \frac{1}{2}\lala L^2\,\rara + \frac{1}{6}\lala L^3\, \rara + \ldots\right) =
 \\
= \lala\ L\ \rara + \frac{1}{2}\Big(\lala L^2\ \rara - \lala\ L\, \rara\,^2\Big)
+ \frac{1}{6}\Big(\lala L^3\, \rara - 3\,\lala L^2\,\rara\ \lala\ L\ \rara
+ 2\, \lala\ L\ \rara\,^3\Big) + \ldots
= \sum_k \frac{1}{k!}\blabla L^k\brabra_{conn}
\ee
to the r.h.s. of (\ref{Bdet}), one concludes that
$\log B_5(x|z)$ can be represented as a sum of connected correlators.
In this case $e^L = \det (x-M)$ and
\be
L = \Tr \log (x-M) \equiv \sum_i \log (x-U_i) =
\sum_i \int^x\! \frac{d{\tilde x}}{{\tilde x}-U_i} =
\int^x \Tr \frac{d{\tilde x}}{{\tilde x}-M}
\label{Lmult}
\ee
Thus
\be
\blabla L^k \brabra_{conn} = \int^x\ldots \int^x
\blabla \Tr\frac{dx_1}{x_1-M}\ \ \ldots\ \ \Tr\frac{dx_k}{x_k-M}\brabra_{conn}
= \int^x\ldots \int^x
\rho_k(x_1,\ldots,x_k)
\ee
is a $k$-fold integral of a $k$-fold connected multi-resolvent
$\rho_k(x_1,\ldots, x_k)$ for the
beta-ensemble (\ref{Bdet}). One obtains therefore
\be\label{WyK5}
\boxed{
\log {B_5(x|q)\over B_4(q)} =
{1\over 2b^2}\int^x W'(x')dx'+
\sum_k \frac{1}{k!} {\int^x\ }^{\otimes k}
\rho_k(x_1,\ldots, x_k)}
\ee
Similarly, for the $n$-point conformal block we get from (\ref{Bdetn})
\be\label{WyKn}
\boxed{
\log {B_n(x_a|q)\over B_4(q)} = \sum_a{1\over 2b^2}\int^{x_a} W'({\tilde x})d{\tilde x}+
\sum_k \frac{1}{k!} \sum_{a_1,\ldots,a_k=1}^{n-4}
\int^{x_{a_1}}\ldots\int^{x_{a_k}}
\rho_k({\tilde x}_{a_1},\ldots, {\tilde x}_{a_k})}
\ee
As explained in detail in \cite{AMMEO},
the multi-resolvents are poly-differentials on spectral curve,
recursively defined from the Virasoro-like constraints
(the Ward identities for the matrix model or beta-ensemble).
This construction
(sometimes called as "topological recursion")
depends only on spectral curve
with distinguished coordinates endowed with a generating differential.
In the multi-Penner model with the potential
$W(x)$
the spectral curve is
$\beta^2y^2 = W'(x)^2 + \sum_i \frac{\beta c_i}{x-z_i}$
with the coefficients $c_i$ being the linear combinations of the $N$-variables
with $\alpha$-dependent coefficients
(for particular examples of multi-resolvents in this case
see \cite{DV,Ito,Egu,Wilmm,Kanno}).
From now on, by making a shift, we absorb the term $W'(x)$ into the definition of
one-point $\rho_1(x)$ (which is very natural as well within the
framework of \cite{AMMEO}).

Formulas (\ref{WyK5}) and (\ref{WyKn}) contains the {\it exact} multi-resolvents,
including contributions of all genera,
\be
\rho_k = \sum_{p\geq 0} \hbar^{p-1}\rho^{(p|k)}
\ee
They coincide with the expressions conjectured in \cite{Wylsurf,Kanno}
(and prove them) for all values of $\beta=b^2$, not only for $\beta=1$ and $Q=0$.
In general case one has just to consider
the beta-ensemble multi-resolvents instead of the matrix model ones,
exploited in \cite{Wylsurf,Kanno}.

\paragraph{Comment.} {\it One has to be very careful with fixing the values of four
external dimensions, corresponding to the beta-ensemble producing the resolvents.
Indeed, for fixed $\alpha_1$, $\alpha_2$ and
$\alpha_3$, the fourth charge $\alpha_4$ is determined by size of the beta-ensemble
(or numbers of the screening operators inserted) and the number of degenerated
fields inserted, (\ref{consl}).
For instance, the value of $\alpha_4$ used in $B_5(x|q)$
differs by $1/2b$ from that in $B_4(z)$, i.e. the average in formula (\ref{Bdet})
is calculated with the beta-ensemble corresponding to $\alpha_4$ shifted
by $1/2b$ as compared to that, describing the Nekrasov function itself.
However, in planar limit, the difference disappears.}

\bigskip

As suggested in \cite{MMS1,MMS}, being based on the matrix model
experience \cite{DVph,MMSW}, the free energy
can be viewed as a (double-deformed) prepotential with $\rho_1$
playing the role of the generating differential:
\be
\label{SWMM2}
\frac{\p \log B_4(q)}{\p a_I} = b^2\oint_{B_I} \rho_1(x), \\
a_I = \oint_{A_I} \rho_1(x)
\ee
This conjecture still remains to be proved.
In sect.\ref{SWfro} we consider a {\it weaker} form of
this conjecture, in the limits of small $\epsilon_2$.

\section{SW theory from the limit $\epsilon_2\to 0$
\label{SWfro}}

\subsection{The limit of the conformal block}

Now we are going to consider the limit of $\epsilon_2\to 0$. To do this,
we restore the parameters
$\epsilon_{1,2}$ of deformation of the Nekrasov functions in the conformal
blocks by rescaling the charges
$\alpha\to \alpha/ g_s $ (i.e. the potential $W(x)\to W(x)/ g_s $ \rf{betapot})
with the string coupling $g_s^2\equiv -\epsilon_1\epsilon_2$.
In the limit of small $\epsilon_2$
in the beta-ensemble with $\beta=b^2=-\epsilon_1/\epsilon_2$
\be
B_4\sim\int \prod_i dx_i \exp\left({1\over g_s}W(x_i)\right)\prod_{i<j}(x_i-x_j)^{2\beta}
\ee
the multi-resolvents behave as
\be\label{rhol}
\rho_k\sim g_s^{2k-2}
\ee
(as an illustration, we list in the Appendix
first few multi-resolvents in the simplest Gaussian case).
It means that when $\epsilon_2\to 0$ and
therefore $g_s\to 0$, only the one-point resolvent
survives in (\ref{WyK5}).

As we already noted above using the knowledge from matrix models \cite{DVph,MMSW}
one can calculate the matrix model partition function using the spectral curve,
endowed with a generating differential. We expect the same claim to be correct for the
beta-ensembles, and since the partition function is now given by $B_4(q)$,
formula (\ref{SWMM2}) should be valid,
where, as usual, the integrals are taken over the $A$- and $B$-cycles of the
spectral curve determined by planar limit of the one-point resolvent.

Now taking the limit $\epsilon_2\to 0$, using (\ref{WyK5})
and (\ref{rhol}) and noting that, in this limit,
$\log B_4$ behaves as $1/g_s^2$ \cite{NF} one finally obtains
\be\label{B5}
B_5(x|q) = \exp \left(-\frac{1}{\epsilon_1\epsilon_2}F(\epsilon_1)
+ \frac{1}{\epsilon_1}S(x;\epsilon_1) + O(\epsilon_2)\right)
\ee
where $F(\epsilon_1)$ does not depend on $x$, while it can and does
depend on $q$, $dS(x;\epsilon_1)\equiv \epsilon_1 \rho_1$ and
\be\label{76}
a = \oint_{A} dS(x;\epsilon_1), \\
\frac{\p F(\epsilon_1)}{\p a} = \oint_{B} dS(x;\epsilon_1)
\ee
where we have rescaled $a\to a/\epsilon_1$ as compared with formula
(\ref{SWMM2}).

\subsection{The Schr\"odinger equation for $S(x)$}

Now one can obtain the Schr\"odinger equation for the ratio of the
conformal blocks
\be
\label{psibb}
{B_5(x|q)\over B_4(q)}=\exp\left({S(x;\epsilon_1)\over \epsilon_1}\right)
\equiv \psi(x)
\ee
To do this, consider solution to the equation
\be\label{85}
\left( b^2\p^2_x + {2x-1\over x(x-1)}\p_x+
{\cal O}\right) B_5(x|q) = 0
\ee
where the operator
\be\label{Gaudin}
{\cal O} =
- \frac{q(q-1)}{x(x-1)(x-q)}\p_q + \frac{1}{\epsilon_1\epsilon_2}{\cal V}(x|z)
=- \frac{q(q-1)}{x(x-1)(x-q)}\p_q+\\
+ {1\over\epsilon_1\epsilon_2}{1\over x(1-x)}\left[\Delta_{1/2b}
+ \frac{\Delta_1}{x} - \frac{\Delta_2}{x-1} - \Delta_3
+ \frac{q^2-(2q-1)x}{(x-q)^2}\Delta_4\right]
\ee
acts only on the $q$-variable.
Then in the leading order in $\epsilon_2^{-1}$ one has \cite{MT}
\be
\left(\frac{\p S}{\p x}\right)^2 + \epsilon_1\frac{\p^2 S}{\p x^2}
= \frac{q(q-1)}{x(x-1)(x-q)}\p_q F(\epsilon_1) + {\cal V}(x|q)
\ee
or
\be\label{ShE}
\boxed{
\Big(-\epsilon_1^2 \p_x^2 + {\cal V}(x)\Big)\psi(x) = \frac{(q-1)E}{x(x-1)(x-q)}
\psi(x)}
\ee
where
$E$ is an $x$-independent quantity
\be
E = {\p F(\epsilon_1)\over\p\log q}
\ee
Note that the limit $\epsilon_2\to 0$ in (\ref{85}) is quite unusual: in such a limit
$b = -\epsilon_1/\epsilon_2 \rightarrow \infty$,
unlike the naive semiclassical limit of the Schr\"odinger equation
(where $b$ would rather go to zero).
Instead in this limit $V_{1/2b} \rightarrow V_0 = 1$
and $B_5(x|q) \rightarrow B_4(q) \ \stackrel{AGT}{=}
\exp\left(-\frac{\F(\epsilon_1,\epsilon_2)}
{\epsilon_1\epsilon_2}\right)$.
Only after one picks up the $\epsilon_2^{-2}$ terms in
the equation, they combine into a Schr\"odinger-like equation
with $\epsilon_1$, playing the role of the Planck constant,
and the semiclassical expansion in small $\epsilon_1$ can
be considered, along the lines of \cite{MMbz}.

\subsection{Examples of different gauge theories}

Thus, we have established that the monodromies
of the wavefunction of the
Sch\"odinger equation (\ref{ShE}) with the Plank constant $\hbar\equiv\epsilon_1$,
i.e. $\oint dS=\oint d\log\psi(x)/\epsilon_1$ are described (\ref{76}) by the
YY function $F(\epsilon_1)$ (i.e. by the Nekrasov function $\F(\epsilon_1,\epsilon_2)$
at $\epsilon_2\to 0$). In particular, the quantization condition of this
Schr\"odinger equation implies that the B-period, which is nothing but the
Bohr/Sommerfeld integral equals $2\pi \hbar (n+1/2)$.

This construction was first discussed for the periodic Toda case
(=pure gauge theory) in \cite{MMbz}. In the $SU(2)$ case one can
easily reproduce the corresponding construction from
sect.~\ref{nonconf}, the potential in the Shr\"odinger equation from
(\ref{nonconf2}) is just $V(x)=\Lambda^2\cosh x$, therefore
\rf{psibb} satisfies \be\label{ShET} \Big(-\epsilon_1^2 \p_x^2 +
\Lambda^2\cosh x\Big)\psi(x) = E\psi(x) \ee More examples were
considered in \cite{MMS1} and in \cite{MT} (further details will
appear in \cite{GZE}). In particular, the case of the gauge theory
with adjoint matter hypermultiplet with mass $m$, which is described
by the Calogero model. In the $SU(2)$ case it is obtained from
eq.(\ref{34}) and leads to the Schr\"odinger equation with elliptic
potential \be\label{ShEC} \Big(-\epsilon_1^2 \p_x^2 +
m(m-\epsilon_1)\wp (x)\Big)\psi(x) = E\psi(x) \ee Of course,
equations (\ref{76}) become really restrictive in the $SL(N)$ case
with $N>2$, when there many $A_I$- and $B_I$-cycles and many periods
$a_I$. However, this case is related to conformal blocks of $W_N$
algebras \cite{Wagt,MMMM}. Analysis of surface operators in these
models can also be easily performed, but this is beyond the scope of
the present paper (see recent papers \cite{Wyllast,Yam} devoted to
this case).

Still, one has to expect that the whole construction of this section is directly
generalized. Indeed, it was proposed and partly checked in \cite{MMbz,MMS1}, that
in the $SU(N)$ case, the role of the Schr\"odinger equation is played by
the Fourier transform of the Baxter equation for the corresponding integrable system.
For instance, the pure gauge $SU(N)$ theory is described by the periodic Toda chain
on $N$ sites, and the corresponding Baxter equation is given by
\be
P_N(\lambda)Q(\lambda)=Q(\lambda+i\hbar)+Q(\lambda-i\hbar)
\ee
where $P_N(\lambda)$ is a polynomial of degree $N$ with coefficients being
the conserved quantities, and $Q(\lambda)$ is the Baxter $Q$-operator.
Thus, the corresponding Schr\"odinger equation is of the form
\be
\left[P_N\left(i\hbar{\p\over\p x}\right)+\cosh x\right]\psi(x)=0
\ee
At $N=2$
$P_2(\lambda)=(\lambda^2-E)/\Lambda^2$ and one obtains
(\ref{ShET}).

The Calogero case is more involved, however, there is also the equation in the
separated variables in this case, which can be considered as the substitute of
(\ref{ShE}), see, for instance, for $N=3$ \cite[eq.(55)]{MMS1}.

However, the most intriguing is the case of the theory with fundamental matter
hypermultiplets with masses $m_a$. As expected from SW theory, this case is
described by the (non-compact) $sl(2)$ (XXX) chain \cite{intSWd}.
The Baxter equation in this case is \cite{MMS1}
\be
P_N(\lambda)Q(\lambda)=K_+(\lambda)Q(\lambda+i\hbar)+K_-(\lambda)Q(\lambda-i\hbar)
\ee
where $K_{\pm}(\lambda)=\prod_a^{N_{\pm}}\left(\lambda -m_a\right)$ and
$N_+ + N_-=N_f$ is the number of matter hypermultiplets (the answer does not depend
on how one parts these hypermultiplets into two sets $N_+$ and $N_-$).
Note that in the case of $N=2$ one does not come to the Schr\"odinger equation
\rf{Gaudin}
of the previous subsection. However, the checks of the first terms (in particular, those
done in \cite{MT}) shows that the both equations lead to the same result! It means
that the Gaudin magnet which, corresponding to
(\ref{Gaudin}), gives rise to the same results as the XXX-chain, at least, in the case of
$N=2$. This point definitely deserves further investigation.

\subsection{Perturbative limit of gauge theories}

This construction, obtained proved indirectly from the
beta-ensemble representation of the conformal block, can be also tested
immediately for the first terms in $\hbar$ and $\Lambda$. It has been done for
various cases in \cite{MMbz,MT}. Note, however, that, for the
perturbative contribution, i.e. in the leading order in $\Lambda$ it can be
checked exactly in $\hbar$. Indeed, let us first look at eq.(\ref{ShET}): its perturbative
limit is described by the Liouville equation \cite{intSWd} (one
has first to shift $x\to x-2\log\Lambda$ and then consider small $\Lambda$ in
eq.(\ref{ShET}), then only one of the exponents remains):
\be\label{ShEL}
\left(-\epsilon_1^2 \p_x^2 + \Lambda^2\exp\left( -x\right)\right)\psi(x) = E\psi(x)
\ee
In this limit, the $A$- and $B$-cycles are degenerate. Note that the cycles
in (\ref{76}) and the corresponding curve are determined
completely by the semiclassical limit of the Schr\"odinger equation,
i.e. by corresponding Seiberg-Witten curve, which becomes rational in this limit.
In particular, the $A$-cycle degenerates into a pair of marked points on the curve,
while the $B$-cycle extends
from the one turning point $x_c$, $E=\Lambda^2\exp(-x_c)$ to infinity
(encircling them), and the
corresponding monodromy of the wave-function is determined by logarithm of
the ratio of asymptotics at infinity:
\be\label{90}
\psi (x)\stackreb{x\to\infty}{\longrightarrow}-{\pi\over\sin{2\pi\lambda\over
\epsilon_1}}
\left[{1\over\Gamma (1+{2\lambda\over \epsilon_1})}\left({\Lambda \over
\epsilon_1}
\right)^{2\lambda/\epsilon_1}e^{x\lambda/\epsilon_1}-
{1\over\Gamma (1-{2\lambda\over \epsilon_1})}\left({\Lambda
\over \epsilon_1}
\right)^{-2\lambda/\epsilon_1} e^{-x\lambda\epsilon_1}
\right]\\
{1\over\epsilon_1}\oint_B dS=\oint_B {\p\log\psi\over\p x}=
\log{c_+(\lambda)\over c_-(\lambda)}
\ee
Here $\lambda\equiv\sqrt{-E}=a$ is pure imaginary and
\be\label{cpm}
c_\pm (\lambda)=
\pm{1\over\Gamma (1\pm {2\lambda\over \epsilon_1})}
\left({\Lambda \over
\epsilon_1}
\right)^{\pm2\lambda/\epsilon_1}
\ee
Formula (\ref{90}) coincides with the perturbative expression for the derivative of
the YY function w.r.t. $a$ and the quantization
condition imposed on the Bohr-Sommerfeld integral
\be
\log{c_+(\lambda)\over c_-(\lambda)}=2\pi in,\ \ \ \ n\in \mathbb{Z}
\ee
coincides with \cite[eq.(6.6)]{NS}.

In fact, the functions $c_\pm(\lambda)$ are proportional to the
Harish-Chandra functions which determine the Plancherel measure on the set of
irreducible unitary representations contributing to the Whittaker model.
Moreover, the $S$-matrix in the integrable system is determined by the Harish-Chandra
functions, see further details and references in \cite{GKMMMO}.

Thus, it is clear, that our consideration
can be easily generalized to generic situation, and
the perturbative result is still determined by logarithm of the ratio of two
asymptotics, i.e. by ratio of two Harish-Chandra functions.
Indeed, for instance, the perturbative limit of the
$SL(N)$-Toda case is described by the conformal (non-periodic) Toda system
\cite{intSWd}, and the corresponding Harish-Chandra functions are \cite{GKMMMO}
\be
c_w(\vec\lambda)\sim \prod_{\vec\alpha\in\Delta_+}{1\over\Gamma (1 - {w(\vec\lambda)\cdot
\vec\alpha\over \epsilon_1})}
\ee
with $w$ being an element of the Weyl group, and
$\Delta_+$ here is the set of all positive roots.
Choosing the basis  (for the $sl(N)$ algebra) $\vec\lambda\cdot\vec\alpha=a_i-a_j$
for all
$i,j=1,\ldots,N$, $i<j$, one easily gets the proper ratios of the Harish-Chandra functions:
\be
\boxed{
{c_{i,+}(\vec\lambda)\over c_{i,-}(\vec\lambda)}=-\left({\Lambda\over\epsilon_1}\right)^{2Na_i\over\epsilon_1}
\prod_{j\neq i}{\Gamma (1- {a_i-a_j\over \epsilon_1})
\over\Gamma (1+ {a_i-a_j\over \epsilon_1})}}
\ee
for all $i=1,\ldots,N$. This immediately leads to the quantization conditions,
coinciding with those of \cite[eq.(6.6)]{NS}.

Finally, consider the case of gauge theory with adjoint matter, described by
the Calogero model (and restrict it here only for the $N=2$ case, i.e. equation (\ref{ShEC})).
The perturbative limit is given by trigonometric Calogero-Moser-Sutherland
model \cite{intSWd}, i.e. by equation
\be
\left(-\epsilon_1^2 \p_x^2 + {m(m-\epsilon_1)\over \sinh^2 x}\right)\psi(x) = E\psi(x)
\ee
The solution to this equation has asymptotics ($\lambda^2=-E$, i.e.
$\lambda$ is again pure imaginary)
\be
\psi(x)\stackrel{x\to\infty}{\sim}\ {\pi\over \sin\pi\left({\lambda\over\epsilon_1}+
{m\over\epsilon_1}\right)}
{e^{-x\lambda/\epsilon_1}\over\Gamma \left(-{\lambda\over\epsilon_1}\right)
\Gamma \left({m\over\epsilon_1}+{\lambda\over\epsilon_1}\right)}
+{\pi\over \sin \pi{m\over\epsilon_1}}
{e^{x\lambda/\epsilon_1}\over \Gamma \left({\lambda\over\epsilon_1}\right)
\Gamma \left({m\over\epsilon_1}-{\lambda\over\epsilon_1}\right)
}
\ee
i.e. the Harish-Chandra functions are
\be
c_\pm(\lambda)\sim {1\over\Gamma \left(\pm{\lambda\over\epsilon_1}\right)
\Gamma \left({m\over\epsilon_1}\mp{\lambda\over\epsilon_1}\right)}
\ee
Logarithm of their ratio again equals to the derivative of the perturbative
part of the YY function and the corresponding quantization
condition coincides with \cite[eq.(6.9)]{NS}.

\section{Conclusion}

In this paper we collected some knowledge about
the degenerate conformal blocks and their
possible application to the study of AGT relations.
The main application so far
is that insertion of the degenerate primary
and appropriate restriction of the additional
intermediate dimension
converts the conformal block into a "wave function",
which, in the limit $\epsilon_2\to 0$ provides the
Seiberg-Witten representation for the one-parameter deformed
prepotential $F(\epsilon_1) = \left.\F\right|_{\epsilon_{2}=0}$, playing also
the role of the YY function.
The differential equation for the degenerate conformal block
turns into a Shr\"odinger-like equation,
which can be also related to the Baxter quantization of the
spectral curve, arising in the SW representation
of the original prepotential $F_{SW} =\left.\F\right|_{\epsilon_{1,2}=0}$.

Thus, despite this ``wave function'' itself (i.e. degenerate conformal block)
is perfectly well-defined for the double-epsilon deformation,
when both $\epsilon_1$ and $\epsilon_2$
are non-vanishing, its interpretation is found only in the
limit of $\epsilon_2\to 0$. Let us remind here, that
Nekrasov function $\F = \F(\epsilon_1,\epsilon_2)$
is a double-deformation of
the original Seiberg-Witten prepotential $F_{SW}$ in two directions:
introducing non-vanishing string coupling
$g_s = \sqrt{-\epsilon_1\epsilon_2}\neq 0$ \cite{LMN}
and the non-vanishing screening charge
$b = \sqrt{-\epsilon_1/\epsilon_2}\neq 1$.
In the
matrix model approach to AGT relations \cite{MMSproof}:
the string coupling governs the topological recursion \cite{AMMEO},
while the second deformation turns matrix model into the
beta-ensemble with $\beta = b^2$.
Understanding is, unfortunately, much worse if the double-deformation
is taken symmetrically for both non-vanishing
$\epsilon_{1,2}\neq 0$, which is
more natural in the original definition of Nekrasov functions \cite{LMNS}.

These two deformations are also different from the point of view of an
integrable systems \cite{SW1,RG,intSWd}: the first, corresponding to $\epsilon_1\neq 0$,
looks like being equivalent to a standard quantum-mechanical quantization of a classical
integrable system, while the second, associated with $\epsilon_2\neq 0$ is rather switching
on the flows of the quasiclassical hierarchy \cite{KriW} (in already quantized problem!).
In particular, turning on the non-vanishing $\epsilon_2$
adds the non-stationary term $\p/\p\log\Lambda$ to the stationary
Schr\"odinger equation (\ref{nonconf2}), and this is exactly the
deformation of the original integrable system in the
direction of the  time-variable $\log\Lambda$ \cite{RG,MN}.
It would be extremely important to extend the role of
degenerate conformal blocks
(or surface operator insertions on the other side of
the AGT relation) to the better understanding of the second
deformation for $\epsilon_2\neq 0$.

\bigskip

Also interesting, though much more straightforward,
is the extension of above discussion from the four-point
Virasoro conformal blocks on sphere (and one-point conformal block on torus)
to generic situation with
arbitrary number of punctures, arbitrary genus and, moreover, for
arbitrary chiral algebras.
In all these cases the surface operator insertions
provide an exhaustive description of the $\epsilon_1\neq 0$
deformation.
All of them are well defined also for $\epsilon_2\neq 0$,
but their role in description of the doubly deformed
prepotentials $\F(\epsilon_1,\epsilon_2)$ still remains to be
revealed.
In all these cases the discrete $\epsilon_1\leftrightarrow \epsilon_2$
symmetry is explicitly broken by the choice of particular
degenerate primary, used for the insertions.

\section*{Acknowledgements}

Our work is partly supported by Russian Federal Nuclear Energy Agency,
supported by Ministry of Education and Science of the Russian Federation
under contract 02.740.11.0608, by RFBR grants 08-01-00667 (A.Mar.) 10-02-00509-a
(A.Mir.), and 10-02-00499 (A.Mor.), by joint grants 09-02-90493-Ukr,
09-02-93105-CNRSL, 09-01-92440-CE, 09-02-91005-ANF, 10-02-92109-Yaf-a. The
work of A.Mar. has been also supported by the grant of support of Scientific schools LSS-1615.2008.2 and by the Max Planck Society. A.Mar. is grateful to the Max Planck
Institute for Mathematics in Bonn, where this work was finished, for the warm hospitality.

\section*{Appendix}

For illustrative purposes we list here the first few multi-resolvents in the Gaussian
$\beta$-ensemble \cite{Gbeta}:
\be
\rho_1(x)=N\left({1\over x}+g_s^2{1+(N-1)\beta\over x^3}+
g_s^4{3-5\beta+3\beta^2+5\beta N-5\beta^2 N+2N^2\beta^2\over x^5}
+\ldots\right)\rightarrow\\
\stackrel{\epsilon_2\to\infty}{\longrightarrow}N\left({1\over x}+{(N-1)\over x^3}\epsilon_1^2+
{2N^2-5N+3\over x^5}\epsilon_1^4+{5N^3-22N^2+32N-15\over x^7}\epsilon_1^6
+\ldots\right)
\\
\\
\rho_2(x,y)=Ng_s^2\left({1\over x^2y^2}+g_s^2
{3-3\beta+3N\beta\over x^4y^2}+g_s^2{3-3\beta+3N\beta\over x^2y^4}
+2g_s^2{1-\beta+\beta N\over x^3y^3}+\ldots\right)
\longrightarrow\\
\stackrel{\epsilon_2\to\infty}{\longrightarrow}Ng_s^2\left({1\over x^2y^2}+
\epsilon_1^2 (N-1)\left\{{3\over x^2y^4}+{2\over x^3y^3}+{3\over x^4y^2}\right\}+
\epsilon_1^4\left\{(15-25N+10N^2)\left({1\over x^6y^2}+{1\over x^2y^6}\right)+\right.
\right.\\
+\left.\left.{15-27N+12N^2\over x^4y^4}+{12-20N+8N^2\over x^5y^3}+
{12-20N+8N^2\over x^3y^5}
\right\}
\ldots\right)
\\
\\
\rho_3(x,y,z)={2Ng_s^4\over x^2y^2z^2}\left(
{1\over x}+{1\over y}+{1\over z}+2g_s^2(1-\beta+\beta N)
\left\{{3\over x^3}+{3\over y^3}+{3\over z^3}+{2\over xyz}+\right.\right.\\+
\left.\left. {3\over x^2y}+{3\over xy^2}+{3\over x^2z}+{3\over xz^2}+
{3\over y^2z}+{3\over yz^2}\right\}+\dots
\right)
\\
\\
\rho_4(x,y,z,w)={2Ng_s^6\over x^2y^2z^2w^2}\left({3\over x^2}+
{3\over y^2}+{3\over z^2}+{3\over w^2}+{4\over xy}+{4\over xz}+{4\over xw}+{4\over yz}
+{4\over yw}+{4\over zw}+\ldots
\right)
\ee
In the paper we use the resolvents of the DF $\beta$-ensemble and do not need
the Gaussian model resolvents themselves.
However, the Gaussian resolvents, which are much simpler,
show nevertheless an important property true' for any potential: the
resolvents $\rho_k$ behaves as $g_s^{2k-2}$ as $\epsilon_2\to 0$ (still being functions of
$\epsilon_1$). This implies that
the naturally normalized quantities would be $\beta^k\rho_k$ rather than $\rho_k$.
Exactly these quantities enter the topological recursion (loop equations).
Therefore, these $\beta^k\rho_k$ are the quantities which remain finite in the limit
of $\beta\to\infty$, while $\rho_k/\rho_1\to\beta^{1-k}\to 0$ for $k\ge 2$, which was
used in (\ref{WyK5}).

Note also that, after $\epsilon_2$ is put equal to zero, there still exists a genus counting,
however, the t'Hooft limit now corresponds to the double scaling limit
of $N\to\infty$ and $N\epsilon_1^2=$fixed.


\begin{thebibliography}{12}

\bibitem{AGT} L.Alday, D.Gaiotto and Y.Tachikawa,
Lett.Math.Phys. {\bf 91} (2010) 167-197, arXiv:0906.3219

\bibitem{Wagt} N.Wyllard,
%\emph{$A_{N-1}$ conformal Toda field theory correlation functions from
%conformal $N=2$ $SU(N)$ quiver gauge theories},
JHEP {\bf 0911} (2009) 002, arXiv:0907.2189;\\
A.Mironov and A.Morozov, Nucl.Phys. {\bf B825} (2009) 1-37, arXiv:0908.2569

\bibitem{3}
N.Drukker, D.Morrison and T.Okuda, JHEP {\bf 0909} (2009) 031, arXiv:0907.2593;\\
A.Marshakov, A.Mironov and A.Morozov, arXiv:0907.3946; JHEP 11 (2009) 048, arXiv:0909.3338;\\
S.Iguri and C.Nunez, JHEP {\bf 11} (2009) 090 , arXiv:0908.3460;\\
D.Nanopoulos and D.Xie, arXiv:0908.4409;
%\emph{Hitchin Equation, Singularity, and N=2 Superconformal Field Theories},
JHEP {\bf 1003} (2010) 043, arXiv:0911.1990;
%{\it Hitchin Equation, Irregular Singularity, and $N=2$ Asymptotical Free Theories},
arXiv:1005.1350;
%{\it $N=2$ Generalized Superconformal Quiver Gauge Theory},
arXiv:1006.3486;\\
N.Drukker, J.Gomis, T.Okuda and J.Teschner,
JHEP {\bf 1002} (2010) 057, arXiv:0909.1105;\\
A.Mironov and A.Morozov, Phys.Lett. {\bf B682} (2009) 118-124, arXiv:0909.3531;\\
A.Gadde, E.Pomoni, L.Rastelli and S.Razamat, JHEP {\bf 1003} (2010) 032, arXiv:0910.2225;\\
L.Alday, F.Benini and Y.Tachikawa, Phys.Rev.Lett. {\bf 105} (2010) 141601, arXiv:0909.4776;\\
S.Kanno, Y.Matsuo, S.Shiba and Y.Tachikawa,
Phys.Rev. {\bf D81} (2010) 046004, arXiv:0911.4787;\\
R.Poghossian,
%\emph{Recursion relations in CFT and N=2 SYM theory},
JHEP {\bf 0912} (2009) 038,  arXiv:0909.3412;\\
G.Bonelli and A.Tanzini,
%\emph{Hitchin systems, N=2 gauge theories and W-gravity},
arXiv:0909.4031;\\
J.-F.Wu and Y.Zhou,
%\emph{From Liouville to Chern-Simons, Alternative Realization
%of Wilson Loop Operators in AGT Duality},
arXiv:0911.1922;\\
L.Hadasz, Z.Jaskolski and P.Suchanek,
%\emph{Recursive representation of the torus 1-point conformal block},
arXiv:0911.2353;
%{\it Proving the AGT relation for N_f = 0,1,2 antifundamentals},
arXiv:1004.1841;\\
G.Giribet,
%\emph{On triality in $N=2$ SCFT with $N_f=4$},
 JHEP {\bf 01} (2010) 097, arXiv:0912.1930;\\
V.Alba and And.Morozov,
%\emph{Check of AGT Relation for Conformal Blocks on Sphere},
Nucl.Phys. {\bf B840} (2010) 441-468, arXiv:0912.2535;\\
M.Fujita, Y.Hatsuda, Y.Koyama and T.-Sh.Tai,
%\emph{Genus-one correction to asymptotically
%free Seiberg-Witten prepotential from Dijkgraaf-Vafa matrix model},
JHEP {\bf 1003} (2010) 046, arXiv:0912.2988;\\
M.Taki,
%\emph{On AGT Conjecture for Pure Super Yang-Mills and W-algebra},
arXiv:0912.4789;
%{\it Surface Operator, Bubbling Calabi-Yau and AGT Relation},
arXiv:1007.2524;\\
Piotr Sulkowski,
%\emph{Matrix models for $\beta$-ensembles from Nekrasov partition functions},
JHEP {\bf 1004} (2010) 063, arXiv:0912.5476;\\
N.Nekrasov and E.Witten, arXiv:1002.0888;\\
R.Santachiara and A.Tanzini,
%\emph{Moore-Read Fractional Quantum Hall wavefunctions and SU(2) quiver gauge theories},
arXiv:1002.5017;\\
S.Yanagida, arXiv:1003.1049;
arXiv:1010.0528;\\
N.Drukker, D.Gaiotto and J.Gomis
%\emph{The Virtue of Defects in 4D Gauge Theories and 2D CFTs},
arXiv:1003.1112;\\
F.Passerini,
%\emph{Gauge Theory Wilson Loops and Conformal Toda Field Theory},
JHEP {\bf 1003} (2010) 125, arXiv:1003.1151;\\
H.Itoyama and T.Oota,
%\emph{Method of Generationg q-Expansion Coefficients for Conformal Block
%and ${\cal N}=2$ Nekrasov Function by $\beta$-Deformed Matrix Model},
arXiv:1003.2929;\\
Wei He and Yan-Gang Miao,
%{\it Magnetic expansion of Nekrasov theory: the SU(2) pure gauge theory},
arXiv:1006.1214;\\
S.Kanno, Y.Matsuo and S.Shiba,
%{\it Analysis of correlation functions in Toda theory and AGT-W
%relation for SU(3) quiver},
arXiv:1007.0601;\\
C.Kozcaz, S.Pasquetti, F.Passerini and N.Wyllard,
%{\it Affine sl(N) conformal blocks from N=2 SU(N) gauge theories},
arXiv:1008.1412;\\
H.Itoyama, T.Oota and N.Yonezawa,
%{\it Massive Scaling Limit of beta-Deformed Matrix Model of Selberg Type},
arXiv:1008.1861;\\
Ta-Sheng Tai,
%{\it Triality in SU(2) Seiberg-Witten theory and Gauss hypergeometric function},
arXiv:1006.0471;
%{\it Uniformization, Calogero-Moser/Heun duality and Sutherland/bubbling pants},
arXiv:1008.4332;\\
M.Billo, L.Gallot, A.Lerda and I.Pesando,
%{\it F-theoretic vs microscopic description of a conformal N=2 SYM theory},
arXiv:1008.5240;\\
K.Maruyoshi and F.Yagi,
%{\it Seiberg-Witten curve via generalized matrix model },
arXiv:1009.5553;\\
A.Brini, M.Marino and S.Stevan,
%{\it The uses of the refined matrix model recursion},
arXiv:1010.1210;\\
A.Mironov, A.Morozov and A.Shakirov,
%{\it On "Dotsenko-Fateev" representation of the toric conformal blocks},
arXiv:1010.1734;
%{\it Brezin-Gross-Witten model as "pure gauge" limit of Selberg integrals},
arXiv:1011.3481

\bibitem{MMMM}
Andrey Mironov, Sergey Mironov, Alexei Morozov
and Andrey Morozov, arXiv:0908.2064

\bibitem{MMnf}
A.Mironov and A.Morozov,
Phys.Lett. {\bf B680} (2009) 188-194, arXiv:0908.2190

\bibitem{nonconf}
D.Gaiotto, arXiv:0908.0307;\\
A.Marshakov, A.Mironov and A.Morozov,
Phys.Lett. {\bf B682} (2009) 125-129, arXiv:0909.2052;\\
V.Alba and And.Morozov,
%\emph{Non-conformal limit of AGT relation from the 1-point torus conformal block},
JETP Lett. {\bf 90} (2009) 708-712 , arXiv:0911.0363

\bibitem{AGTguk}
L.Alday, D.Gaiotto, S.Gukov, Y.Tachikawa and H.Verlinde,
JHEP {\bf 1001} (2010) 113, arXiv:0909.0945

\bibitem{DV}
R.Dijkgraaf and C.Vafa, arXiv:0909.2453

\bibitem{5dJ}
H.Awata and Y.Yamada, JHEP {\bf 1001} (2010) 125, arXiv:0910.4431;
%{\it Five-dimensional AGT Relation and the Deformed beta-ensemble},
arXiv:1004.5122

\bibitem{NS} N.Nekrasov and S.Shatashvili, arXiv:0908.4052

\bibitem{MMbz} A.Mironov and A.Morozov, %Bohr-Sommerfeld,
%\emph{Nekrasov Functions and Exact Bohr-Zommerfeld Integrals},
JHEP {\bf 04} (2010) 040, arXiv:0910.5670;
%\emph{Nekrasov Functions from Exact BS Periods: the Case of SU(N)},
J.Phys. {\bf A43} (2010) 195401, arXiv:0911.2396
\\
A.Popolitov, arXiv:1001.1407

\bibitem{Ito}
H.Itoyama, K.Maruyoshi and T.Oota,
%\emph{Notes on the Quiver Matrix Model and 2d-4d Conformal Connection},
Prog.Theor.Phys. {\bf 123} (2010) 957-987, arXiv:0911.4244

\bibitem{Egu} T.Eguchi and K.Maruyoshi,
%\emph{Penner Type Matrix Model and Seiberg-Witten Theory},
arXiv:0911.4797;
%{\it Seiberg-Witten theory, matrix model and AGT relation},
arXiv:1006.0828

\bibitem{Wilmm} R.Schiappa and N.Wyllard,
%\emph{An $A_r$ threesome: Matrix models, $2d$ CFTs and $4d$ N=2 gauge theories},
arXiv:0911.5337

\bibitem{MMS1} A.Mironov, A.Morozov and Sh.Shakirov,
%\emph{Matrix Model Conjecture for Exact BS Periods and Nekrasov Functions},
JHEP {\bf 02} (2010) 030, arXiv:0911.5721

\bibitem{FLit1} V.Fateev and I.Litvinov, JHEP {\bf 1002} (2010) 014, arXiv:0912.0504

\bibitem{MMS} A.Mironov, A.Morozov and Sh.Shakirov,
%\emph{Conformal blocks as Dotsenko-Fateev Integral Discriminants},
arXiv:1001.0563\\
A.Mironov, A.Morozov and And.Morozov,
%{\it Matrix model version of AGT conjecture and generalized Selberg integrals},
arXiv:1003.5752

\bibitem{Wylsurf}
C.Kozcaz, S.Pasquetti and N.Wyllard,
%{\it A & B model approaches to surface operators and Toda theories},
arXiv:1004.2025

\bibitem{MT} K.Maruyoshi and M.Taki,
%{\it Deformed Prepotential, Quantum Integrable System and Liouville Field Theory},
arXiv:1006.4505

\bibitem{Kanno} H.Awata, H.Fuji, H.Kanno, M.Manabe and Y.Yamada,
%{\it Localization with a Surface Operator, Irregular Conformal
%Blocks and Open Topological String},
arXiv:1008.0574

\bibitem{CDV} M.C.N.Cheng, R.Dijkgraaf and C.Vafa,
%{\it Non-Perturbative Topological Strings And Conformal Blocks},
arXiv:1010.4573

\bibitem{Wyllast} N.Wyllard,
%{\it W-algebras and surface operators in N=2 gauge theories},
arXiv:1011.0289

\bibitem{Yam} Y.Yamada,
%{\it A quantum isomonodromy equation and its application to N=2 SU(N)
%gauge theories},
arXiv:1011.0292

\bibitem{CFT}
A.Belavin, A.Polyakov, A.Zamolodchikov, Nucl.Phys. {\bf B241} (1984) 333-380;\\
A.Zamolodchikov and Al.Zamolodchikov,
\emph{Conformal field theory and critical phenomena in 2d systems}, 2009 (in Russian)

\bibitem{beta}
P.Di Francesco, M.Gaudin, C.Itzykson and F.Lesage,
Int.J.Mod.Phys. {\bf A9} (1994) 4257-4352, hep-th/9401163\\
A.Zabrodin, arXiv:0907.4929\\
Sh. Shakirov,
%\emph{Exact solution for mean energy of 2d Dyson gas at $\beta = 1$},
arXiv:0912.5520;\\
L.Chekhov,
%{\it Logarithmic potential beta-ensembles and Feynman graphs },
arXiv:1009.5940

\bibitem{confmamo} A.Marshakov, A.Mironov, and A.Morozov,
%\emph{Generalized matrix models as conformal field theories: Discrete case},
Phys.Lett. {\bf B265} (1991) 99\\
S.Kharchev, A.Marshakov, A.Mironov, A.Morozov and S.Pakuliak,
%\emph{Conformal Matrix Models as an Alternative to Conventional Multi-Matrix Models},
Nucl.Phys. {\bf B404} (1993) 17-750,  arXiv:hep-th/9208044

\bibitem{DVmamo}
K.Demeterfi, N.Deo, S.Jain and C.-I Tan, Phys.Rev. {\bf D42} (1990) 4105-4122\\
J.Jurkiewicz,
%{\it Regularization of one-matrix models},
Phys.Lett. {\bf 245} (1990) 178\\
\u{C}.Crnkovi\'c and G.Moore, Phys.Lett. {\bf B257} (1991)
322\\
G.Akemann and J.Ambj{\o}rn,
%{\it New universal spectral correlators},
J.Phys. {\bf A29} (1996) L555--L560, cond-mat/9606129\\
G.Akemann,
%{\it Higher genus correlators for the Hermitian matrix model
%with multiple cuts},
Nucl.Phys. {\bf B482} (1996) 403, hep-th/9606004\\
G.Bonnet, F.David, and B.Eynard,
%{Breakdown of universality in multi-cut matrix models},
J.Phys. {\bf A33} (2000) 6739--6768

\bibitem{DVph} R.Dijkgraaf and C.Vafa,
  Nucl.Phys. {\bf B644} (2002) 3, hep-th/0206255;
  Nucl.Phys. {\bf B644} (2002) 21, hep-th/0207106;
hep-th/0208048

\bibitem{DVnext}
H.Itoyama and A.Morozov,
Nucl.Phys.B657:53-78,2003, hep-th/0211245;
Phys.Lett. B555 (2003) 287-295, hep-th/0211259;
Prog.Theor.Phys. 109 (2003) 433-463, hep-th/0212032;
Int.J.Mod.Phys. A18 (2003) 5889-5906, hep-th/0301136
\\
A.Klemm, M.Marino and S.Theisen,
JHEP 0303 (2003) 051, hep-th/0211216

\bibitem{LMNS}
G.Moore, N.Nekrasov, S.Shatashvili, Nucl.Phys. {\bf B534} (1998) 549-611, hep-th/9711108;
hep-th/9801061\\
A.Losev, N.Nekrasov and S.Shatashvili, Commun.Math.Phys. {\bf 209} (2000) 97-121, hep-th/9712241;
ibid. 77-95, hep-th/9803265

\bibitem{NF} N.Nekrasov, Adv.Theor.Math.Phys. {\bf 7} (2004) 831-864,
hep-th/0206161

\bibitem{SW} N.Seiberg and E.Witten, Nucl.Phys., {\bf B426} (1994) 19-52,
hep-th/9408099; Nucl.Phys., {\bf B431} (1994) 484-550,
hep-th/9407087

\bibitem{SW1} A.Gorsky, I.Krichever, A.Marshakov, A.Mironov, A.Morozov,
Phys.Lett., {\bf B355} (1995) 466-477, hep-th/9505035

\bibitem{intSWd}
E.Martinec,
%Integrable structures in supersymmetric gauge and string theories,
Phys.Lett., {\bf B367} (1996) 91-96;\\
T.~Nakatsu and K.~Takasaki,
  %``Whitham-Toda hierarchy and N = 2 supersymmetric Yang-Mills theory,''
  Mod.\ Phys.\ Lett.\  A {\bf 11} (1996) 157
  [arXiv:hep-th/9509162];\\
R.Donagi and E.Witten, Nucl.Phys., {\bf B460} (1996) 299-334,
hep-th/9510101;\\
A.Gorsky, A.Marshakov, A.Mironov and A.Morozov,
%N=2 SQCD and integrable spin chains; rational case $N_{f}\le 2N_{c}$,
Phys.Lett., {\bf B380} (1996) 75-80, hep-th/9603140; hep-th/9604078;\\
A.Gorsky, S.Gukov and A.Mironov,
%Supersymmetric Yang-Mills
%theories, integrable systems and their stringy/brane origin-I,
%hep-th/9707120 to appear in Nucl.Phys.B
Nucl.Phys., {\bf B517} (1998) 409-461, hep-th/9707120;
%Supersymmetric Yang-Mills
%theories, integrable systems and their stringy/brane origin-II,
%hep-th 9710239 to appear in Nucl.Phys.B
Nucl.Phys., {\bf B518} (1998) 689, 9710239;\\
H.W.Braden, A.Marshakov, A.Mironov and A.Morozov, Nucl.Phys., {\bf
B573} (2000) 553 hep-th/9906240; Phys.Lett.,
 {\bf B448} (1999) 195, hep-th/9812078;
Nucl.Phys., {\bf B573} (2000) 553, hep-th/9906240;\\
A.Gorsky and A.Mironov, Nucl.Phys.,
{\bf B550} (1999) 513, hep-th/9902030; hep-th/0011197\\
A.Mironov and A.Morozov, hep-th/0001168

\bibitem{intSW}
E.Martinec and N.Warner,
%Integrable systems  and supersymmetric
%Yang-Mills theory,
Nucl.Phys., {\bf 459} (1996) 97;\\
A.Gorsky, A.Marshakov,
%Towards effective topological gauge theories on the spectral curves,
Phys.Lett., {\bf B374} (1996) 218-224;\\
H.Itoyama and A.Morozov,
%Integrability and Seiberg-Witten theory; curves and periods,
Nucl.Phys., {\bf B477} (1996) 855-877, hep-th/9511126;
Nucl.Phys., {\bf B491} (1997) 529-573, hep-th/9512161, hep-th/9601168;\\
E.D'Hoker, I.M.Krichever and D.H.Phong,
Nucl.Phys., {\bf B489} (1997) 179-210;
Nucl.Phys., {\bf B489} (1997) 211-222;\\
N.Nekrasov,
%Five dimensional gauge theories and relativistic integrable systems,
Nucl.Phys., {\bf B531} (1998) 323-344, hep-th/9609219;\\
A.Marshakov, A.Mironov,
%Prepotentials in 5d and 6d theories from integrable systems, hep-th 9711239
Nucl.Phys., {\bf B518} (1998) 59-91;\\
A.Marshakov, {\sl Seiberg-Witten Theory and Integrable Systems},
World Scientific, Singapore, 1999\\
A.Mironov and A.Morozov, Phys.Lett., {\bf B475} (2000) 71, arXiv:hep-th/9912088;\\
H.Braden and A.Marshakov, Nucl.Phys. {\bf B595} (2001) 417-466; hep-th/0009060;\\
N.Nekrasov and S.Shatashvili,
%{\it Supersymmetric vacua and Bethe ansatz},
Nucl.Phys. Proc.Suppl. {\bf B192-193} (2009) 91-112, arXiv:0901.4744;
%{\it Quantum integrability and supersymmetric vacua},
arXiv:0901.4748

\bibitem{dua} D.Gaiotto,
%{\it N=2 dualities},
arXiv:0904.2715\\
D.Gaiotto, G.W.Moore and A.Neitzke,
%{\it Wall-crossing, Hitchin Systems, and the WKB Approximation},
arXiv:0907.3987

\bibitem{6dth} E.Witten,
%Solution of N=2 supersymmetric theories via M theory,
Nucl.Phys., {\bf B500} (1997) 3-42, hep-th/9703166;\\
A.Marshakov, M.Martellini, A.Morozov,
%{\it Insights and Puzzles from Branes: 4d SUSY Yang-Mills from 6d Models},
Phys.Lett. {\bf B418} (1998) 294-302,  hep-th/9706050;\\
E.Witten,
%{\it Geometric Langlands From Six Dimensions},
 arXiv:0905.4795

\bibitem{ZFLON}
V.Fateev, A.Litvinov, A.Neveu and E.Onofri,
%{\it Differential equation for four-point correlation function in Liouville field
%theory and elliptic four-point conformal blocks},
J.Phys. {\bf A42} (2009) 304011, arXiv:0905.2280\\
The parametrization used in this paper was first proposed in:\\
Al.B. Zamolodchikov,
%{\it Two-dimensional conformal symmetry and critical 4-spin correlation-functions
%in the Ashkin-Teller model,
Sov.Phys. JETP {\bf 63(5)} (1986) 1061-1066

\bibitem{Brav} A.Braverman,
%{\it Instanton counting via affine Lie algebras I: Equivariant J-functions of (affine)
%flag manifolds and Whittaker vectors},
arXiv:math/0401409\\
A.Braverman and P.Etingof,
%{\it Instanton counting via affine Lie algebras II: from Whittaker vectors to the
%Seiberg-Witten prepotential},
arXiv:math/0409441

\bibitem{AMMEO}
A.Alexandrov, A.Mironov and A.Morozov,
%\emph{Partition functions of matrix models as the first special functions
%of string theory. I: Finite size Hermitean 1-matrix model},
Int.J.Mod.Phys. {\bf A19} (2004) 4127, hep-th/0310113;
Teor.Mat.Fiz. {\bf 150} (2007) 179-192, hep-th/0605171;
Physica {\bf D235} (2007) 126-167, hep-th/0608228; JHEP {\bf 12} (2009) 053,
arXiv:0906.3305;\\
A.Alexandrov, A.Mironov, A.Morozov, P.Putrov,
%{\it Partition Functions of Matrix Models as the First Special Functions of String Theory. II. Kontsevich Model},
Int.J.Mod.Phys. {\bf A24} (2009) 4939-4998, arXiv:0811.2825;\\
B.Eynard,
%\emph{All genus correlation functions for the hermitian 1-matrix model},
JHEP \textbf{0411} (2004) 031, hep-th/0407261;\\
L.Chekhov and B.Eynard,
%\emph{Hermitean matrix model free energy:
%Feynman graph technique for all genera},
JHEP \textbf{0603} (2006) 014, hep-th/0504116;
%\emph{Matrix eigenvalue model: Feynman graph technique for all genera},
JHEP \textbf{0612} (2006) 026, math-ph/0604014; \\
N.Orantin,
%\emph{Symplectic invariants, Virasoro constraints
%and Givental decomposition},
arXiv:0808.0635;\\
I.Kostov and N.Orantin, arXiv:1006.2028;\\
L.Chekhov, B.Eynard and O.Marchal,
%{\it Topological expansion of beta-ensemble model and quantum algebraic geometry in the sectorwise approach},
arXiv:1009.6007

\bibitem{KT}
K.Kozlowski and J.Teschner, arXiv:1006.2906.

\bibitem{EO} T.Eguchi and H.Ooguri,
%{\it Conformal and Current Algebras on a General Riemann Surface},
Nucl.Phys. {\bf B282} (1987) 308-328

\bibitem{Ino}
V.Inozemtsev, Comm.Math.Phys., {\bf 121} (1989) 629

\bibitem{FreeF} Vl.Dotsenko and V.Fateev, Nucl.Phys. {\bf B240} (1984) 312-348

\bibitem{Bos}
A.Gerasimov, A.Marshakov, A.Morozov, M.Olshanetsky, S. Shatashvili,
%\emph{Wess-Zumino-Witten model as a theory of free fields},
Int.J.Mod.Phys. {\bf A5} (1990) 2495-2589;\\
A.Gerasimov, A.Marshakov and A.Morozov,
%\emph{ Free Field Representation Of Parafermions And Related Coset Models},
Nucl.Phys. {\bf B328} (1989) 664, Theor.Math.Phys. {\bf 83} (1990)
466-473;
%\emph{Hamiltonian Reduction Of Wess-Zumino-Witten Theory
%From The Point Of View Of Bosonization},
Phys.Lett. {\bf B236} (1990) 269, Sov.J.Nucl.Phys. {\bf 51} (1990)
371-372

\bibitem{MMSW} L.Chekhov and A.Mironov,
  Phys.Lett. {\bf B552} (2003) 293, hep-th/0209085;\\
   V.~Kazakov and A.~Marshakov,
  %``Complex curve of the two matrix model and its tau-function,''
  J.\ Phys.\ A  {\bf 36} (2003) 3107
  [arXiv:hep-th/0211236];\\
L.~Chekhov, A.~Marshakov, A.~Mironov and D.~Vasiliev,
  %``DV and WDVV,''
  Phys.\ Lett.\  B {\bf 562} (2003) 323
  [arXiv:hep-th/0301071];
Proc. Steklov Inst.Math. {\bf 251} (2005) 254,
hep-th/0506075\\
A.Alexandrov, A.Mironov and A.Morozov,
%\emph{Unified description of correlators in non-Gaussian phases of Hermitean matrix model},
Int.J.Mod.Phys. \textbf{A21} (2006) 2481-2518, hep-th/0412099;
%\emph{Solving Virasoro Constraints in Matrix Models},
Fortsch.Phys. \textbf{53} (2005) 512-521, hep-th/0412205;\\
A.Mironov, Theor.Math.Phys. \textbf{146} (2006) 63-72,
hep-th/0506158;\\
A.~Marshakov,
  %``Matrix models, complex geometry and integrable systems. I,''
  Theor.\ Math.\ Phys.\  {\bf 147} (2006) 583
  [Teor.\ Mat.\ Fiz.\  {\bf 147} (2006) 163]
  [arXiv:hep-th/0601212];Theor.\ Math.\ Phys.\  {\bf 147} (2006) 777
  [Teor.\ Mat.\ Fiz.\  {\bf 147} (2006) 399]
  [arXiv:hep-th/0601214];\\
B.Eynard and N.Orantin,
%{\it Invariants of algebraic curves and topological expansion},
arXiv:math-ph/0702045

\bibitem{GKMMMO} A.Gerasimov, S.Kharchev, A.Marshakov, A.Mironov, A.Morozov and
M.Olshanetsky,
%{\it Liouville Type Models in Group Theory Framework. I. Finite-Dimensional Algebras},
Int.J.Mod.Phys. {\bf A12} (1997) 2523-2584, hep-th/9601161

\bibitem{GZE} L.Grechishnikov and E.Timiriasov, to appear\\
E.Zenkevich, to appear

\bibitem{LMN}
  A.~S.~Losev, A.~Marshakov and N.~Nekrasov,
  %``Small instantons, little strings and free fermions,''
  in Ian Kogan memorial volume
  {\em From fields to strings:
circumnavigating theoretical physics}, 581-621
  [arXiv: hep-th/0302191].

\bibitem{MMSproof} A.Mironov, A.Morozov and Sh.Shakirov,
{\it Towards a proof of AGT conjecture by methods of matrix models}, to appear

\bibitem{KriW}
I.~Krichever,
%``The tau-function of the universal Whitham hierarchy,
%matrix models and topological field theories'',
Commun. Pure. Appl. Math. {\bf 47} (1992) 437
[arXiv: hep-th/9205110].

\bibitem{RG}
M.Matone, Phys.Lett., {\bf B357} (1995) 342;
hep-th/9506102;
Phys.Rev., {\bf D53} (1996) 7354\\
A.Gorsky, A.Marshakov, A.Mironov and A.Morozov, Nucl.Phys., B527 (1998) 690-716, hep-th/9802004 \\
J.Edelstein, M.Gomez-Reino, M.Marino and J.Mas, Nucl.Phys., B574 (2000) 587, hep-th/9911115 \\
J.Edelstein, M.Gomez-Reino and J.Mas, Nucl.Phys., B561 (1999) 273, hep-th/9904087.

\bibitem{MN}
A.~Marshakov and N.~Nekrasov,
  %``Extended Seiberg-Witten theory and integrable hierarchy,''
  JHEP {\bf 0701} (2007) 104
  [arXiv: hep-th/0612019];\\
A.~Marshakov,
  %``On Microscopic Origin of Integrability in Seiberg-Witten Theory,''
  Theor.\ Math.\ Phys.\  {\bf 154} (2008) 362
  arXiv:0706.2857 [hep-th].

\bibitem{Gbeta} A.Morozov and Sh.Shakirov,
%{\it The matrix model version of AGT conjecture and CIV-DV prepotential},
arXiv:1004.2917

\end{thebibliography}
\end{document}